\documentclass[final,journal]{IEEEtran}
\IEEEoverridecommandlockouts
\usepackage{cite}
\usepackage{amsmath,amssymb,amsfonts}
\usepackage{algorithmic}
\usepackage{graphicx}
\usepackage{textcomp}
\usepackage{xcolor}
\usepackage{bm,soul}
\usepackage{comment}
\usepackage{tikz}
\usepackage{pgfplots}
\usepackage{float}
\usepackage{color}
\usepackage{setspace}
\usepackage{scalerel}
\usepackage{mathtools}
\usepackage[acronym,shortcuts]{glossaries}
\usepackage{todonotes,url}
\usepackage{bbm}
\usepackage{balance}

\def\BibTeX{{\rm B\kern-.05em{\sc i\kern-.025em b}\kern-.08em
    T\kern-.1667em\lower.7ex\hbox{E}\kern-.125emX}}
\def\overlinebold#1{\ThisStyle{\ooalign{%
  $\SavedStyle\mkern3mu\overline{\phantom{\mathrm{#1}}}$\cr $\SavedStyle\bm #1$}}}
  
\newacronym[plural=AoAs,firstplural=angles of arrival (AoAs)]{aoa}{AoA}{angle-of-arrival}
\newacronym[plural=AoDs,firstplural=angles of departure (AoDs)]{aod}{AoD}{angle-of-departure}
\newacronym[plural=CFRs,firstplural= channel frequency responses (CFRs)]{CFR}{CFR}{channel frequency response}
\newacronym[plural=CIRs,firstplural= channel impulse responses (CIRs)]{CIR}{CIR}{channel impulse response}
\newacronym[plural=RNNs,firstplural= recurrent neural networks (RNNs)]{RNN}{RNN}{recurrent neural network}
\newacronym{$k$-NN}{$k$-NN}{k-Nearest Neighbors}
\newacronym{awgn}{AWGN}{additive white Gaussian noise}
\newacronym{BS}{BS}{base station}
\newacronym{br}{BR}{best response}
\newacronym{CDF}{CDF}{cumulative distribution function}
\newacronym{ce}{CE}{channel estimation}
\newacronym{cearc}{CEARC}{channel estimation with adaptive RIS configuration}
\newacronym{cr-pla}{CR-PLA}{challenge-response PLA}
\newacronym{CSI}{CSI}{channel state information}
\newacronym{det}{DET}{detection error tradeoff}
\newacronym{dft}{DFT}{discrete Fourier transform}
\newacronym{FA}{FA}{false alarm}
\newacronym{lmmse}{LMMSE}{linear minimum mean square error}
\newacronym{LOS}{LOS}{line of sight}
\newacronym{LRT}{LRT}{likelihood ratio test} 
\newacronym{ls}{LS}{least-square}
\newacronym{LSTM}{LSTM}{long short-term memory}
\newacronym{LT}{LT}{likelihood test}
\newacronym{MAP}{MAP}{maximum a posteriori probability}
\newacronym{mbce}{MBCE}{MUSIC-based channel estimation}
\newacronym{mdp}{MDP}{misdetection probability}
\newacronym{MIMO}{MIMO}{multiple-input multiple-output}
\newacronym{miso}{MISO}{multiple-input single-output}
\newacronym{ML}{ML}{machine learning}
\newacronym{mmwave}{mmWave}{millimeter-wave}
\newacronym{mrc}{MRC}{maximum ratio combining}
\newacronym{mse}{MSE}{mean squared error}
\newacronym{music}{MUSIC}{multiple signal classification}
\newacronym{ofdm}{OFDM}{orthogonal frequency division multiplexing}
\newacronym{omp}{OMP}{orthogonal matching pursuit}
\newacronym{parafac}{PARAFAC}{parallel factor}
\newacronym{pdf}{pdf}{probability density function}
\newacronym{PLA}{PLA}{physical layer authentication}
\newacronym{pmf}{PMF}{probability mass function}
\newacronym{RIS}{RIS}{reconfigurable intelligent surface}
\newacronym{simo}{SIMO}{single-input multiple-output}
\newacronym{snr}{SNR}{signal-to-noise ratio}
\newacronym{ue}{UE}{user equipment}
\newacronym{ULA}{ULA}{uniform linear array}
\newacronym{zf}{ZF}{zero-forcing}

\begin{document}
\sloppy
\title{Game-Theoretic Analysis of RIS-Assisted Challenge-Response PLA -- \\ Part II: Partial Channel Knowledge
\thanks{This work was partially supported by European Union (EU) COST Action CA22168—Physical Layer Security for Trustworthy and Resilient 6G Systems (6G-PHYSEC). }
}

\author{Anna V. Guglielmi$^{1}$, Linda Senigagliesi$^{2}$, Marco Baldi$^{3}$, and Stefano Tomasin$^{1}$ \\
\IEEEauthorblockA{$^{1}$University of Padova, Italy,
  $^{2}$ETIS UMR 8051, CYU, ENSEA, CNRS, Cergy, France,\\
 $^{3}$Università Politecnica delle Marche, Ancona, Italy\\
	 email: annavaleria.guglielmi@unipd.it, linda.senigagliesi@ensea.fr,  m.baldi@univpm.it, stefano.tomasin@unipd.it
       }
       }

\maketitle

\begin{abstract}
We consider a physical layer authentication scenario in which a legitimate user (Alice) communicates with an authenticator (Bob) through a reconfigurable intelligent surface (RIS), controlled by Bob. An attacker (Trudy) aims to impersonate Alice by suitably precoding her messages and still transmitting them through the RIS. This two-part study analyzes this attack-defense interaction between Bob and Trudy under perfect (Part I) and partial (Part II) channel knowledge by all the agents. Part I formulates the joint RIS selection and precoder design under perfect channel knowledge as a finite zero-sum game, establishing exact channel subspace conditions for attack indistinguishability alongside equilibrium mixed strategies. Part II extends the defense to incomplete channel information, where each agent observes only its own channel, by formulating a finite-action Bayesian game and characterizing the resulting Bayesian Nash equilibrium. Simulations in geometric and 3GPP TR 38.901 UMi-NLOS channels reveal that authentication performance is governed by available spatial degrees of freedom, channel correlation, and operating signal-to-noise ratio. The proposed equilibrium strategies effectively balance communication rate and security, outperforming conventional benchmark policies.
\end{abstract}

\section{Introduction}

Authentication is the process enabling communicating entities to mutually verify their identities or allow one party to verify another. Asymmetric cryptographic primitives under computational security represent the standard solution; however, advances in cryptanalysis and quantum computing necessitate post-quantum alternatives whose complexity can constrain resource-limited devices~\cite{Lopez2025}.

Physical Layer Authentication (\ac{PLA}) complements cryptography by operating outside the computational security paradigm. \ac{PLA} has been investigated across diverse environments, including \ac{ofdm} and \ac{MIMO} communications~\cite{Baracca2012, hou2014physical} and underwater acoustic channels~\cite{diamant2019cooperative}. Decision mechanisms span from classical Neyman-Pearson statistical tests~\cite{maurer2000authentication} to modern machine learning models~\cite{Fang2018}. These schemes extract physical signal features, such as \ac{CFR} and \ac{CIR}~\cite{11003929}, as well as \ac{aoa}~\cite{Pham2023,Srinivasan2024}, to differentiate legitimate transmitters from adversaries. Both classical statistical tools and machine learning classifiers are applied for user discrimination~\cite{Senigagliesi2021}. In standard tag-based \ac{PLA}, the observed channel state during transmission is evaluated against a reference fingerprint acquired during a trusted association phase.

Reconfigurable Intelligent Surfaces (\Acp{RIS}) dynamically shape propagation environments and enhance \ac{PLA} effectiveness. Optimizing \ac{RIS} reflections toward target locations increases channel distinctiveness and improves tag-based authentication accuracy~\cite{Meng2023}, including non-line-of-sight vehicular scenarios~\cite{Shawky2023}. Channel-estimating hybrid \ac{RIS} architectures enable \ac{CFR}-based authentication~\cite{Selim2023}, whereas other approaches combine \ac{RIS} spatial control with pre-shared cryptographic keys~\cite{Zhang2023} or employ convolutional neural networks to withstand time-varying channels~\cite{Liu2024}.

Beyond static enhancement, \ac{RIS} reconfigurability enables challenge-response PLA (\ac{cr-pla})~\cite{Tomasin2022, tomasin2024analysis, icc2, guglielmi2026multi}. In \ac{cr-pla}, the verifier randomly selects an \ac{RIS} configuration as a challenge and checks whether the resulting cascaded response matches expected legitimate propagation. Security depends heavily on attacker knowledge: when the verifier randomly perturbs the \ac{RIS} configuration around its optimal communication state, an adversary forging responses via a \ac{MAP} channel estimate can degrade \ac{cr-pla} into tag-based PLA if equipped with sufficient spatial degrees of freedom and perfect channel knowledge~\cite{tomasin2024analysis}. Optimizing \ac{RIS} phase perturbation distributions maximizes rate bounds under misdetection constraints across single- and multi-antenna setups in Rayleigh fading~\cite{icc2, guglielmi2026multi}, while geometric analyses establish conditions under which a fully informed multi-antenna attacker becomes indistinguishable from a single-antenna legitimate user~\cite{Senigagliesi25}.

Game theory provides a natural framework for jointly modeling attack and defense strategies as mutual best responses. Although game-theoretic PLA models exist~\cite{Xiao2017GT,Zhou2022GT}, early formulations lacked reconfigurable environments or challenge-response dynamics. In Part I~\cite{partI}, \ac{cr-pla} was modeled as a finite zero-sum game under perfect channel knowledge. However, perfect knowledge represents a worst-case assumption rarely met in practice, as legitimate channels are securely established while verifiers typically lack prior knowledge of attacker positions and channel states.

In this two-part work, we consider a RIS-assisted \ac{cr-pla} system in which a multi-antenna verifier, Bob, controls a \ac{RIS} to authenticate a multi-antenna legitimate user, Alice, while a multi-antenna attacker, Trudy, precodes her transmission to impersonate Alice. All signals reach Bob through the RIS.
In this second part of the work, we consider the scenario in which each player observes only its own channel realization and knows the statistics of the opponent's channel, whose correlation with Alice's channel models the spatial proximity of Trudy to Alice.
Compared to the case of complete knowledge of the channel, discussed in Part I of this work, this incomplete information changes the nature of the problem in three respects. First, each player optimizes an interim utility conditioned on its private information, so that the interaction must be modeled as a Bayesian game~\cite{Harsanyi1967}, whose solution concept is the Bayesian Nash equilibrium (BNE), rather than as a complete-information zero-sum game with a minimax solution. Second, the players' strategies become type-dependent: Bob's randomization policy depends on the realization of Alice's channel, and Trudy's precoding strategy on the realization of her own channel. Third, Trudy can no longer compute a dedicated precoder for each RIS configuration and channel realization, and must rely on a precoder codebook designed offline from channel statistics. To the best of our knowledge, the Bayesian game arising in RIS-assisted \ac{cr-pla} under incomplete channel information has not been investigated so far.

\subsection{Contributions of Part II}

The main contributions of this paper can be summarized as follows:
\begin{itemize}
\item We formulate RIS-assisted \ac{cr-pla} under partial channel knowledge as a finite-action Bayesian game between Bob, who selects the RIS configuration from a finite codebook, and Trudy, who selects her precoder from a finite codebook. The players' types are their privately observed channel realizations, and each player's statistical knowledge of the opponent's channel is captured by a common prior over the types.
Through the Harsanyi transformation, we establish the existence of a BNE in type-dependent mixed strategies.
 
\item We characterize the BNE through type-dependent indifference conditions and support constraints, which yield Bob's randomized RIS policy for each realization of Alice's channel, Trudy's randomized precoding strategy for each realization of her own channel, and the corresponding equilibrium misdetection probability.
 
\item We design Trudy's precoder codebook offline using the Linde–Buzo–Gray (LBG) algorithm, yielding codewords that represent clusters of least-squares precoders computed for representative channel realizations and RIS configurations. We also include in Trudy's codebook, for each of her channel types, a configuration-agnostic precoder that extends the fixed attack in~\cite{tomasin2024analysis} by reproducing Trudy's estimate of Alice's channel, also when Trudy has fewer antennas than RIS elements.
 
\item We evaluate the equilibrium security under both a geometric multipath channel model and the 3GPP TR~38.901 UMi-NLOS model~\cite{3gpp38901}, quantifying the role of the channel correlation, the number of RIS--Bob paths, the number of antennas at Bob and Trudy, and the SNR. We compare the proposed equilibrium policy with deterministic, uniform, and state-of-the-art-inspired baselines. With respect to perfect channel knowledge, incomplete information markedly favors the defender: in the 3GPP scenario, the misdetection probability no longer saturates as Trudy's number of antennas grows, while the SNR and the Alice--Trudy distance become dominant factors. We also show that RIS codebooks obtained by strongly perturbing the communication-optimal configuration outperform DFT-based codebooks, since the precoders of an attacker that cannot adapt to the instantaneous configuration are only loosely tied to the geometry of any single configuration.
\end{itemize}

Unlike~\cite{icc2, guglielmi2026multi}, in this work Bob's policy is optimized with respect to authentication security only, and the rate--security trade-off is assessed a posteriori by comparing the equilibrium misdetection probability obtained with different RIS codebooks. A joint optimization of both objectives is left for future work. 

The rest of the paper is organized as follows. Section~\ref{sec:mod} introduces the system model, including a summary of the \ac{cr-pla} mechanism and the resulting misdetection probability. Section~\ref{sec:probForm} describes the attack and defenses considered in the partial-channel knowledge scenario and models Bob and Trudy's behavior through a game. In Section~\ref{sec:design}, the strategies of the two players are characterized through the Bayesian Nash equilibrium via the indifference principle. Section~\ref{sec:res} presents numerical results under both a geometric multipath channel model and the 3GPP TR~38.901 UMi-NLOS model, and Section~\ref{sec:concs} concludes the paper. 

\glsresetall

\section{System Model}
\label{sec:mod}

\begin{figure}[t!]
    \centering
    \includegraphics[width=90mm]{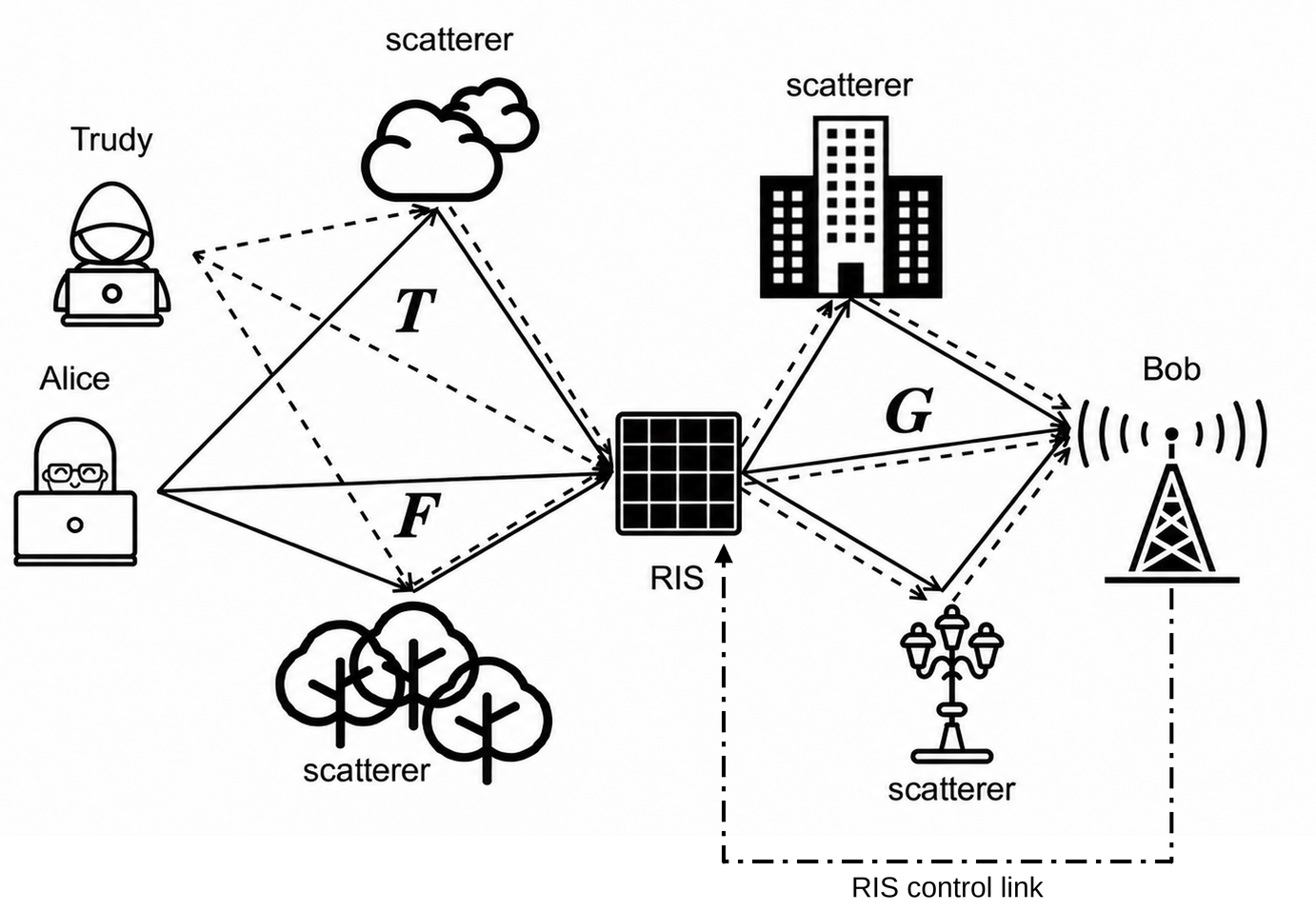}
    \caption{System model.}
    \label{fig:sys_mod}
\end{figure}

We consider the system model shown in Fig.~\ref{fig:sys_mod}, where the \ac{BS} (Bob) aims to authenticate a \ac{ue} (Alice) in a \ac{MIMO} communication scenario. Alice is equipped with \(K\) transmitting antennas, whereas Bob employs a \ac{ULA} with \(M\) antennas. The signal transmitted by Alice reaches Bob through a \ac{RIS}, while a blockage obstructs the direct Alice--Bob link.

An attacker device, referred to as Trudy, attempts to impersonate Alice by transmitting messages that Bob may mistake as originating from the legitimate user. Trudy is equipped with a \ac{ULA} of $N_{\rm T}$ antennas. We assume that no direct communication is possible between Trudy and Bob and that all signals transmitted by Trudy reach Bob through the RIS.
Transmissions occur at \ac{mmwave} frequencies. The antenna elements of the ULAs are uniformly spaced by $d=\lambda_c/2$, where $\lambda_c$ is the carrier wavelength. 

The RIS, controlled by Bob, consists of $N$ reflecting elements spaced by the same distance $d$. Its \(n\)-th element introduces the phase shift
\begin{equation}
\label{phasecons}
    \omega_n=e^{j\varphi_n},
    \qquad n=0,1,\ldots,N-1,
\end{equation}
on the equivalent baseband signal and is assumed to have unitary gain. The RIS configuration matrix is defined as
\begin{equation}
    \bm{\Omega}
    =
    \operatorname{diag}
    \left\{
    \left[e^{j\varphi_0},\ldots,e^{j\varphi_{N-1}}\right]
    \right\}
    =
    \operatorname{diag}\{\bm{\omega}\}.
\end{equation}

We denote the baseband-equivalent Alice--RIS channel by $\bm F\in\mathbb C^{N\times K}$ and the RIS--Bob channel by $\bm G\in\mathbb C^{M\times N}$. The resulting cascaded Alice--RIS--Bob channel is
\begin{equation}
\label{eq:cascadedCh}
    \bm H_{\rm ARB}
    =
    \bm G\bm\Omega\bm F\,.
\end{equation}
Alice transmits suitable pilot symbols that allow Bob to estimate the cascaded channel and use it as a physical layer fingerprint. The pilot signal is assumed to be known to Trudy.

We denote the Trudy--RIS channel by $\bm T\in\mathbb C^{N\times N_{\rm T}}$. Trudy aims to impersonate Alice by producing at Bob a signal as similar as possible to that generated by Alice. To this end, Trudy applies a precoding matrix $\bm Q\in\mathbb C^{N_{\rm T}\times K}$ to the transmitted signal, including the pilot symbols. The resulting cascaded Trudy--RIS--Bob channel is
\begin{equation}
\label{eq:HTR}
    \bm H_{\rm TRB}
    =
    \bm G\bm\Omega\bm T\bm Q
    \in\mathbb C^{M\times K}\,.
\end{equation}

All channel matrices are assumed to remain constant throughout one authentication exchange. Different authentication instances may correspond to different channel realizations drawn according to the statistical model introduced below.
 
We assume here that channels $\bm{T}$ and $\bm{F}$ are statistically dependent, so by the knowledge of $\bm{T}$, some information on $\bm{F}$ is obtained. In Section~\ref{sec:res}, a specific model for the statistical dependence is introduced to obtain numerical results.

\paragraph*{Partial Channel Knowledge Model}

We consider a scenario where Bob and Trudy have asymmetric channel information. Bob knows the realizations of the Alice--RIS channel $\bm F$ and the RIS--Bob channel $\bm G$, but has only statistical knowledge of the Trudy--RIS channel $\bm T$. Conversely, Trudy knows the realizations of $\bm T$ and $\bm G$, but has only statistical knowledge of $\bm F$. Both players know the joint channel distribution. Moreover, Trudy does not observe the RIS configuration selected by Bob, but knows the policy according to which he selects it.

\subsection{Physical Layer Authentication Mechanism}
\label{sec:PLA}
 
We consider the \ac{cr-pla} mechanism of \cite{partI}, which is briefly summarized here. Bob aims to decide whether a received pilot signal has been transmitted by the legitimate user Alice or by the attacker Trudy. The \ac{cr-pla} mechanism consists of two phases. 

In the {\em association phase}, Alice transmits a known pilot matrix, and Bob obtains a reference estimate $\overline{\bm H}(\bm\omega)$ of the legitimate cascaded Alice--RIS--Bob channel. The procedure is repeated for several RIS configurations so that the channel can then be inferred for any other RIS configuration $\bm \omega$, denoted $\overline{\bm H}(\bm\omega)$, \cite{9771077,10625161,piana2026brisc}. This phase is assumed to be protected by higher-layer mechanisms. The association phase must be repeated whenever the legitimate cascaded channel changes significantly.

Then, whenever Bob receives a packet, he performs the {\em verification phase}, i.e., he selects a RIS configuration $\bm\omega$ according to his authentication policy and estimates the channel $\hat{\bm H}$. He then compares $\hat{\bm H}$ with the $\overline{\bm H}(\bm\omega)$
legitimate reference estimate corresponding to the selected RIS configuration. In particular, according to the likelihood test (LT), considering the AWGN noise on the channel estimate,  Bob computes
\begin{equation}
\zeta(\bm\omega,\bm Q)
=
\left\|
\hat{\bm H}-\overline{\bm H}(\bm\omega)
\right\|^2,
\label{eq:zeta}
\end{equation}
and applies the threshold test
\begin{equation}
\begin{cases}
\zeta(\bm\omega,\bm Q) \le \tau,
& \hat{\mathcal H}=\mathcal H_0,\\
\zeta(\bm\omega,\bm Q) > \tau,
& \hat{\mathcal H}=\mathcal H_1,
\end{cases}
\label{eq:thresholdtest}
\end{equation}
where $\tau$ is the authentication threshold.

\paragraph*{Performance Metrics} Two error events may occur. A false alarm occurs when Bob rejects a legitimate transmission from Alice, whereas a misdetection occurs when Bob accepts a forged transmission from Trudy. The corresponding false-alarm probability and \ac{mdp} are
\begin{subequations}
\begin{align}\label{probdef}
P_{\rm FA}
&=
\Pr\left[
\zeta(\bm\omega) > \tau
\,\middle|\,
\mathcal H_0
\right],
\\
P_{\rm MD}
&=
\Pr\left[
\zeta(\bm\omega,\bm Q) \le \tau
\,\middle|\,
\mathcal H_1
\right].
\end{align}
\end{subequations}

For fixed channel realizations $\bm F$ and $\bm T$, a fixed RIS configuration $\bm\omega$, and a fixed attack precoder $\bm Q$, we define the conditional \ac{mdp} as
\begin{equation}\label{eq:pmd_conditional}
P_{\rm MD}(\bm\omega,\bm Q;\bm F,\bm T,\bm G)
=
\Pr\left[
\zeta(\bm\omega,\bm Q) \le \tau
\,\middle|\,
\mathcal H_1,\bm F,\bm T,\bm G,\bm\omega,\bm Q
\right],
\end{equation}
where the probability is taken with respect to the receiver noise and the channel-estimation error. Denoting by $\sigma^2$ the per-entry variance of the estimation error on $\hat{\bm H}-\overline{\bm H}(\bm\omega)$, under $\mathcal H_1$ the normalized statistic $2\zeta(\bm\omega,\bm Q)/\sigma^2$ is noncentral chi-square distributed with $2MK$ degrees of freedom and noncentrality parameter $2\|\bm G\bm\Omega(\bm F-\bm T\bm Q)\|_F^2/\sigma^2$, whereas under $\mathcal H_0$ it is central chi-square distributed with the same degrees of freedom. Hence, $\tau$ is set from the central chi-square distribution to meet the target $P_{\rm FA}$, and~\eqref{eq:pmd_conditional} is computed in closed form through the noncentral chi-square cumulative distribution function. The SNR is defined as $1/\sigma^2$, with all channels normalized to unit average power per entry. Then, if Bob selects the RIS configuration according to a probability distribution $p(\cdot|\bm F)$, the attack success probability of a fixed precoder $\bm Q$ is obtained by averaging the conditional \ac{mdp} over Bob's randomized challenge, namely
\begin{equation}
\overline{P}_{\rm MD}(p, \bm Q; \bm F,\bm T,\bm G)
=
\mathbb E_{\bm\omega\sim p(\cdot|\bm F)}
\left[
P_{\rm MD}(\bm\omega,\bm Q;\bm F,\bm T,\bm G)
\right].
\end{equation}

\section{Attacks and Defenses}\label{sec:probForm}

The \ac{cr-pla} mechanism introduced above makes the \ac{mdp} the natural security metric of the system. In this section, we formulate the corresponding robust design problem, characterizing the fundamental attack and defense objectives induced by the authentication statistic. We then frame it into a game-theory setting, where Bob and Trudy are the players. 

Bob controls the RIS configuration vector $\bm \omega \in \mathbb{T}^N$, where $\mathbb T\triangleq\{z\in\mathbb C:|z|=1\}$, while Trudy controls the precoding matrix $\bm Q \in \mathcal{Q}$, where $\mathcal{Q}=\mathbb C^{N_{\rm T}\times K}$ denotes the continuous admissible precoder space. For fixed channel realizations $(\bm F,\bm T)$, each action pair $(\bm\omega,\bm Q)$ induces the conditional \ac{mdp} defined in \eqref{eq:pmd_conditional}.
Therefore, Bob aims at selecting the RIS configuration, or a randomized RIS configuration policy, so as to reduce this
probability, whereas Trudy aims at selecting the precoder that increases it.

We consider here the general case in which Bob and Trudy have only partial, statistical knowledge of the adversary channel. 
In particular, we assume that Bob and Trudy do not have access to the same channel information, and their strategies can only be designed based on their respective available knowledge. Indeed, Bob knows the instantaneous realization of $\bm F$ and forms his belief about Trudy's channel according to the conditional distribution $p(\bm T\mid\bm F)$. Conversely, Trudy knows the instantaneous realization of $\bm T$ and forms her belief about Alice's channel according to $p(\bm F\mid\bm T)$. Both conditional distributions follow from the
commonly known joint channel model introduced in Section~\ref{sec:mod}. Importantly, while Bob selects $\bm \omega$, Trudy does not observe its instantaneous realization, but only knows the policy according to which Bob selects it. The channel $\bm{G}$ is common knowledge to both players. The case of perfect channel knowledge by Bob and Trudy is analyzed in Part~I of this work,~\cite{partI}.

Under this partial-knowledge model, Bob selects the RIS configuration according
to a randomized policy conditioned on his available information, i.e.,
\begin{equation}
    p_{\bm \omega}(\cdot|\bm{F}) \in \mathcal{P}_{\bm \omega}(\mathbb{T}^N),
    \label{eq:bob_policy}
\end{equation}
where \(\mathcal P_{\bm \omega}(\mathbb T^N)\) denotes the set of probability distributions over the feasible RIS configurations. In turn, Trudy randomly selects the precoder according to a distribution conditioned on her available information, i.e.,
\begin{equation}
   p_{\bm Q}(\cdot|\bm{T}) \in \mathcal{P}_{\bm Q}(\mathcal Q).
    \label{eq:trudy_policy}
\end{equation}
Note that the dependence on the commonly known channel $\bm G$ is omitted from the policy notation for simplicity. 

The overall \ac{mdp} induced by the randomized RIS configuration and precoding policies is obtained by averaging the conditional \ac{mdp} over the joint channel distribution and over the actions selected by the two players, namely
\begin{equation}
\begin{split}
\overline{P}_{\rm MD}
(p_{\bm\omega},p_{\bm Q})
={}&
\mathbb E_{(\bm F,\bm T)\sim p_{\bm F,\bm T}}
\Bigg[
\mathbb E_{\substack{
\bm\omega\sim p_{\bm\omega}(\cdot\mid\bm F)\\
\bm Q\sim p_{\bm Q}(\cdot\mid\bm T)
}}
\left[
P_{\rm MD}(\bm\omega,\bm Q;\bm F,\bm T)
\right]
\Bigg]\,.
\end{split}
\label{eq:attack_prob}
\end{equation}
 
Accordingly, Bob seeks a randomized RIS configuration policy that reduces
the \ac{mdp}, whereas Trudy seeks a randomized precoding policy that increases it. Since directly optimizing these policies over the continuous RIS and precoder spaces is generally intractable, in the following we introduce finite channel and action sets and formulate the resulting strategic interaction.

\subsection{Game Definition}
\label{sec:opt_attack}

Since directly optimizing the policies in~\eqref{eq:bob_policy}--\eqref{eq:trudy_policy} over the continuous RIS and precoder spaces is intractable, due to the non-convexity of $\mathbb{T}^N$, the threshold-based nature of the authentication test, and the coupling between the optimization variables, we introduce the following approximations:
\begin{itemize}
\item The channel realizations defining the players' private information are represented by the finite type sets, i.e., 
\begin{equation}
\bm{F} \in \mathcal F = \{\bm{F}_1, \ldots, \bm{F}_{N_{\mathcal F}}\},
\end{equation}
\begin{equation}
\bm{T} \in \mathcal T = \{\bm{T}_1, \ldots, \bm{T}_{N_{\mathcal T}}\}.
\end{equation}
\item Both players select their actions from finite codebooks, i.e.,
\begin{equation}
\bm \omega \in \mathcal{C}_\Omega = \{ \bm{\omega}_1, \dots, \bm{\omega}_{N_{\Omega}} \}\subseteq \mathbb T^N.
\end{equation}
\begin{equation}
\bm Q \in \mathcal{C}_Q(\bm T_\ell)=\{\bm Q_1,\dots,\bm Q_{N_Q-1},\bm Q_{\mathrm{agn},\ell}\}\subseteq\mathcal{Q},
\end{equation}
where the first $N_Q-1$ precoders are common to all of Trudy’s types, while the last one is the configuration-agnostic precoder, which depends on Trudy’s type $\bm T_\ell$. When the configuration-agnostic precoder is not considered, $\mathcal C_Q$ comprises $N_Q$ codewords common to all types.
\end{itemize}

This yields a game-theoretic framework that preserves the essential strategic structure of the continuous problem.

We frame the optimization problem then as a {\em game} involving two {\em players}: Bob
(defender), who controls the RIS configuration, and Trudy (attacker), who
selects a precoding matrix to forge the cascaded channel.

\paragraph*{Actions}
The sets of actions for the two players are defined by their respective
codebooks $\mathcal C_\Omega$ and $\mathcal C_Q$. In particular, Bob selects a RIS configuration vector \(\bm{\omega}\in\mathcal C_\Omega\), while Trudy selects a precoding matrix from  \(\mathcal C_Q\).

\paragraph*{Payoff Function}
For a given RIS configuration \(\bm{\omega}\), a given precoder $\bm Q$, and fixed channel realizations \((\bm F,\bm T)\), we define Trudy's payoff as the conditional \ac{mdp}, i.e.,
\begin{equation}
\label{eq:payoff}
u(\bm{\omega},\bm Q;\bm F,\bm T)
=
P_{\rm MD}(\bm{\omega},\bm Q;\bm F,\bm T).
\end{equation}
This payoff represents the probability that Bob accepts Trudy's forged
transmission as legitimate. Since Bob aims at rejecting impersonation attempts,
his payoff is \(1-u(\bm{\omega},\bm Q;\bm F,\bm T)\).

\paragraph*{Nature and Incomplete Information}
The asymmetric channel knowledge of Section~\ref{sec:mod} is modeled by letting nature draw the channel pair according to the commonly known joint PMF 
\begin{equation}
\gamma_{k\ell}=\Pr(\bm F=\bm F_k,\bm T=\bm T_\ell).
\end{equation}
The realization of $\bm F$ constitutes Bob's private information, i.e., his \emph{type}, whereas the realization of $\bm T$ constitutes Trudy's type, leading to a \emph{Bayesian game}. After observing their types, the players form their beliefs about the opponent's type as
\begin{equation}\label{eq:beliefs}
\Pr(\bm T_\ell\mid\bm F_k)=\frac{\gamma_{k\ell}}{\sum_{r}\gamma_{kr}},
\qquad
\Pr(\bm F_k\mid\bm T_\ell)=\frac{\gamma_{k\ell}}{\sum_{r}\gamma_{r\ell}}\,.
\end{equation}
The type sets are obtained by drawing $N_{\mathcal F}$ realizations of $\bm F$ from its marginal distribution, so that $\Pr(\bm F_k)=1/N_{\mathcal F}$, and $\gamma_{k\ell}=\Pr(\bm T_\ell\mid\bm F_k)/N_{\mathcal F}$, where $\Pr(\bm T_\ell\mid\bm F_k)$ is obtained from the conditional density $p(\bm T\mid\bm F)$ of the channel model, normalized over $\ell$. Hence, the beliefs in~\eqref{eq:beliefs} depend on each player's own type through the channel correlation; the expressions adopted for the two channel models are given in Section~\ref{sec:VB}.

\section{Strategies Design}
\label{sec:design}
In the finite codebook setting, a Nash equilibrium in pure strategies, i.e., deterministic choices of \(\bm{\omega}\in\mathcal C_\Omega\) for Bob and
\(\bm Q\in\mathcal C_Q\) for Trudy, is not guaranteed to exist: there may be no pair \((\bm{\omega}_i, \bm{Q}_j)\) such that neither player has an incentive to deviate unilaterally. This motivates the use of \emph{mixed strategies},
i.e., probability distributions over the finite action sets, which guarantee equilibrium existence in the resulting finite Bayesian game, while still
including pure strategies as a special case. Due to the incomplete information, the \ac{pmf} used by Bob to choose the RIS configuration is conditioned on his private information \(\bm F\), while the \ac{pmf} used by Trudy to choose the precoder is conditioned on her private information \(\bm T\).

Let \(\Delta^D\) denote the set of all probability distributions over \(D\)
elements, i.e.,
\begin{equation}
\Delta^D =
\left\{
\bm v\in\mathbb R^D:
v_i\geq 0,\ \sum_{i=1}^D v_i=1
\right\}.
\end{equation}
Defining
\begin{equation}
o_i(\bm{F}) = {\mathbb P}[\bm{\omega} = \bm{\omega}_i|\bm{F}]
\end{equation}
\begin{equation}
a_j(\bm{T}) = {\mathbb P}[\bm{Q} = \bm{Q}_j|\bm{T}]\,,
\end{equation}
the mixed strategy of Bob is defined as
\begin{equation}
\bm{o}(\bm F)
=[o_1(\bm{F}), \ldots, o_{N_{\Omega}}(\bm{F})] 
\in \Delta^{N_{\Omega}},
\end{equation}
while a mixed strategy for Trudy is a probability distribution
\begin{equation}
\bm{a}(\bm T)
=[a_1(\bm{T}), \ldots, a_{N_Q}(\bm{T})]
\in \Delta^{N_Q}.
\end{equation}

\paragraph*{Expected Payoff}
 
For given channel realizations \((\bm F,\bm T)\) provided by nature, the {\em expected payoff} of Trudy induced by the mixed-strategy profile is
\begin{equation}
\label{exppayoff}
\begin{split}
&U(\bm o(\bm F),\bm a(\bm T)\mid\bm F,\bm T)
\\
&=
\mathbb E_{\substack{
\bm\omega\sim\bm o(\bm F)\\
\bm Q\sim\bm a(\bm T)
}}
\left[
u(\bm\omega,\bm Q;\bm F,\bm T)
\right]
\\
&=
\sum_{i=1}^{N_\Omega}
\sum_{j=1}^{N_Q}
o_i(\bm F)a_j(\bm T)
u(i,j\mid\bm F,\bm T).
\end{split}
\end{equation}
In \eqref{exppayoff}, we have made explicit the dependence of
the payoff on the channel realization through
\begin{equation}
u(i,j|\bm F,\bm T)
=
P_{\rm MD}(\bm\omega_i,\bm Q_j;\bm F,\bm T).
\end{equation}

By taking the expectation also on the channel, we have
\begin{equation}
\label{exppayoff2}
\begin{split}
\overline U(\bm o,\bm a)
={}&
\mathbb E_{\bm F,\bm T}
\left[
U(\bm o(\bm F),\bm a(\bm T)\mid\bm F,\bm T)
\right]
\\
={}&
\sum_{k=1}^{N_{\mathcal F}}
\sum_{\ell=1}^{N_{\mathcal T}}
\gamma_{k\ell}
\sum_{i=1}^{N_\Omega}
\sum_{j=1}^{N_Q}
o_i(\bm F_k)
a_j(\bm T_\ell)
u(i,j\mid\bm F_k,\bm T_\ell),
\end{split}
\end{equation}
where $\gamma_{k\ell}$ is the commonly known joint PMF of the channel types. Eq.~\eqref{exppayoff2} defines the ex-ante payoff, evaluated before the players observe their private types. After observing their respective types, however, Bob and Trudy optimize their interim expected utilities by averaging over the unknown type of the opponent according to their conditional beliefs. 

\subsection{Optimal Mixed-Strategies}

Both Bob and Trudy maximize their respective utilities by selecting the mixed strategies $\bm o(\bm F)$ and $\bm a(\bm T)$. Bob's utility is the probability of successful discrimination, whereas Trudy's utility is the probability of successful impersonation. After observing their private types,
the players cannot directly use the ex-ante payoff in \eqref{exppayoff2}; instead, each player evaluates an interim expected utility by averaging over the unknown channel of the opponent according to the corresponding conditional belief. 
In particular, the expected payoff for Bob, given his current type \(\bm F\), is defined as the probability of successful discrimination averaged over Trudy's unknown channel
\begin{align}
U_B &(\{\bm o(\bm F), \bm a(\bm T)\}\mid\bm F)
=
\sum_{i=1}^{N_{\Omega}}
\sum_{j=1}^{N_Q}
o_i(\bm{F}) \\ \nonumber
&\qquad \cdot \mathbb E_{\bm T|\bm F}
\left[
a_j(\bm{T})
\left(
1-u(i,j|\bm F,\bm T)
\right)
\right].
\end{align}
The expected payoff for Trudy, given her current type \(\bm T\), is the
probability of successful impersonation averaged over Bob's unknown channel
\begin{align}
U_T &(\{\bm a(\bm T), \bm o(\bm F)\}\mid\bm T) 
=
\sum_{i=1}^{N_{\Omega}}
\sum_{j=1}^{N_Q}
a_j(\bm{T}) \\ \nonumber
&\qquad \cdot \mathbb E_{\bm F|\bm T}
\left[
o_i(\bm{F})
u(i,j|\bm F,\bm T)
\right].
\end{align}
These payoffs reflect the asymmetric view of the players: Bob's strategy $\bm o(\bm F)$ depends on \(\bm F\), while Trudy's strategy $\bm a(\bm T)$ depends on \(\bm T\). Bob and Trudy aim at selecting type-dependent mixed strategies that maximize their respective interim expected utilities. For compactness, we denote the complete mixed strategies by
\begin{equation}
\bm o
=
\{\bm o(\bm F):\bm F\in\mathcal F\},
\qquad
\bm a
=
\{\bm a(\bm T):\bm T\in\mathcal T\}.
\end{equation}

\paragraph*{Bayesian Nash Equilibrium}
A mixed-strategy profile $(\bm o^\star,\bm a^\star)$ is a Bayesian Nash equilibrium (BNE) if, after observing their respective private types, neither player can improve its interim expected utility by unilaterally changing the probability distribution employed for that type. Therefore,
for every type having positive marginal probability,
\begin{equation}
\label{eq:bne1}
\bm o^\star(\bm F)
\in
\arg\max_{\widetilde{\bm o}\in\Delta^{N_\Omega}}
U_B\left(
\widetilde{\bm o},
\bm a^\star
\mid\bm F
\right),
\forall\bm F\in\mathcal F:
\Pr(\bm F)>0,
\end{equation}
\begin{equation}
\label{eq:bne2}
\bm a^\star(\bm T)
\in
\arg\max_{\widetilde{\bm a}\in\Delta^{N_Q}}
U_T\left(
\widetilde{\bm a},
\bm o^\star
\mid\bm T
\right),
\forall\bm T\in\mathcal T:
\Pr(\bm T)>0.
\end{equation}
Here, $\widetilde{\bm o}$ and $\widetilde{\bm a}$ denote unilateral deviations in the action distributions employed by Bob and Trudy, respectively, for the currently observed type. The strategies prescribed for all the other types remain unchanged.
The use of the inclusion symbol in \eqref{eq:bne1} and \eqref{eq:bne2} accounts for the possibility that multiple action distributions yield the same maximum interim utility.

\paragraph*{Existence of a Bayesian Nash equilibrium}
\label{prop:BNE_existence} 

We now prove the existence of a Bayesian Nash equilibrium. By the Harsanyi transformation, the Bayesian game can be represented as a complete-information normal-form game in which the pure strategies of Bob
and Trudy are the contingent mappings $s_B:\mathcal F \rightarrow \mathcal{C}_{\Omega}$ and $s_T:\mathcal T \rightarrow \mathcal{C}_Q$, respectively.
Since both the type spaces and the action spaces are finite, the sets of such contingent strategies are also finite. Note that, although Trudy's action set is type-dependent, its size $N_Q$ is the same for all types. Moreover, the expected utilities induced by the common prior $\{\gamma_{k\ell}\}$ are well-defined and bounded. Therefore, the resulting normal-form game is finite and, by Nash's existence theorem \cite{Nash1950}, admits at least one equilibrium in mixed strategies. 
By the Harsanyi equivalence, such an equilibrium induces a Bayesian Nash equilibrium of the original incomplete-information game~\cite{Harsanyi1968}.  
Since only one type of each player is realized in each instance of the game, randomization over contingent plans can equivalently be represented, in terms of the induced distribution over realized actions and payoffs, through the type-dependent mixed strategies $\bm o^\star(\bm F)$ and
$\bm a^\star(\bm T)$ defined in \eqref{eq:bne1}--\eqref{eq:bne2}.

We denote by
\begin{equation}
V_{\rm MD} = \overline U(\bm o^\star,\bm a^\star)
\label{eq:VMD}
\end{equation}
the equilibrium \ac{mdp}, i.e., the ex-ante payoff~\eqref{exppayoff2} at the BNE. Since the game is constant-sum, $V_{\rm MD}$ is the same for all equilibria.

\subsection{Equilibrium Characterization via Indifference Principle}

Since the channel type spaces and the action spaces are finite, the Bayesian Nash equilibrium can be characterized through the interim payoffs associated with the players' pure actions.

For a given type \(\bm F\in\mathcal F\) having positive marginal probability, let
\begin{equation}
C_i(\bm F;\bm a)
=
\mathbb E_{\bm T|\bm F}
\left[
\sum_{j=1}^{N_Q}
a_j(\bm T)
u(i,j|\bm F,\bm T)
\right],
\,
i=1,\ldots,N_{\Omega},
\label{eq:bob_action_cost}
\end{equation}
denote the expected \ac{mdp} when Bob selects the pure RIS configuration \(\bm\omega_i\), while Trudy follows the type-dependent strategy \(\bm a(\bm T)\). Since Bob aims at minimizing the \ac{mdp}, we consider only the RIS
configurations that are assigned positive probability by his equilibrium
strategy (i.e., $o_i^\star(\bm F)>0$). Each such configuration must yield
the minimum expected \ac{mdp}
\begin{equation}
C_i(\bm F;\bm a^\star)
=
\min_{r=1,\ldots,N_{\Omega}}
C_r(\bm F;\bm a^\star),
\,
\forall i \,\text{such that }o_i^\star(\bm F)>0.
\label{eq:bob_best_response}
\end{equation}

Similarly, for a given type \(\bm T\in\mathcal T\) having positive marginal probability, let
\begin{equation}
D_j(\bm T;\bm o)
=
\mathbb E_{\bm F|\bm T}
\left[
\sum_{i=1}^{N_{\Omega}}
o_i(\bm F)
u(i,j|\bm F,\bm T)
\right],
\,
j=1,\ldots,N_Q,
\label{eq:trudy_action_payoff}
\end{equation}
denote Trudy's expected payoff when she selects the pure precoder \(\bm Q_j\), while Bob follows the type-dependent strategy \(\bm o(\bm F)\). Since Trudy aims at maximizing the misdetection probability, her equilibrium strategy must satisfy 
\begin{equation}
D_j(\bm T;\bm o^\star)
=
\max_{\ell=1,\ldots,N_Q}
D_\ell(\bm T;\bm o^\star), \,
\forall j \, \text{such that }a_j^\star(\bm T)>0.
\label{eq:trudy_best_response}
\end{equation}


Let
\begin{equation}
\mathcal S_B(\bm F)
=
\left\{
i\in\{1,\ldots,N_{\Omega}\}:
o_i^\star(\bm F)>0
\right\}
\end{equation}
and
\begin{equation}
\mathcal S_T(\bm T)
=
\left\{
j\in\{1,\ldots,N_Q\}:
a_j^\star(\bm T)>0
\right\}
\end{equation}
denote the type-dependent supports of Bob's and Trudy's equilibrium strategies, respectively.

According to the indifference principle, for every
$\bm F\in\mathcal F$, all RIS configurations belonging to
$\mathcal S_B(\bm F)$ must yield the same minimum expected
\ac{mdp}, i.e.,
\begin{subequations}\label{eq:bob_support_conditions}
\begin{equation}
C_{i_1}(\bm F;\bm a^\star)
=
C_{i_2}(\bm F;\bm a^\star),
\qquad
\forall i_1,i_2\in\mathcal S_B(\bm F).
\label{eq:bob_indifference}
\end{equation}
The actions outside the support must satisfy
\begin{equation}
C_i(\bm F;\bm a^\star)
\geq
C_r(\bm F;\bm a^\star),
\qquad
\forall i\notin\mathcal S_B(\bm F),\
r\in\mathcal S_B(\bm F).
\label{eq:bob_outside_support}
\end{equation}
\end{subequations}

Likewise, for every $\bm T\in\mathcal T$, all precoders belonging to
$\mathcal S_T(\bm T)$ must yield the same maximum expected payoff, i.e.,
\begin{subequations}\label{eq:trudy_support_conditions}
\begin{equation}
D_{j_1}(\bm T;\bm o^\star)
=
D_{j_2}(\bm T;\bm o^\star),
\qquad
\forall j_1,j_2\in\mathcal S_T(\bm T).
\label{eq:trudy_indifference}
\end{equation}
The actions outside the support must satisfy
\begin{equation}
D_j(\bm T;\bm o^\star)
\leq
D_\ell(\bm T;\bm o^\star),
\qquad
\forall j\notin\mathcal S_T(\bm T),\
\ell\in\mathcal S_T(\bm T).
\label{eq:trudy_outside_support}
\end{equation}
\end{subequations}

The equilibrium probabilities must also satisfy
\begin{subequations}\label{eq:bob_probability_constraints}
\begin{equation}
\sum_{i\in\mathcal S_B(\bm F)}
o_i^\star(\bm F)=1,
\qquad
o_i^\star(\bm F)=0,
\quad
\forall i\notin\mathcal S_B(\bm F),
\label{eq:bob_probability_sum}
\end{equation}
with
\begin{equation}
o_i^\star(\bm F)>0,
\qquad
\forall i\in\mathcal S_B(\bm F).
\label{eq:bob_probability_positive}
\end{equation}
\end{subequations}
and
\begin{subequations}\label{eq:trudy_probability_constraints}
\begin{equation}
\sum_{j\in\mathcal S_T(\bm T)}
a_j^\star(\bm T)=1,
\qquad
a_j^\star(\bm T)=0,
\quad
\forall j\notin\mathcal S_T(\bm T),
\label{eq:trudy_probability_sum}
\end{equation}
with
\begin{equation}
a_j^\star(\bm T)>0,
\qquad
\forall j\in\mathcal S_T(\bm T).
\label{eq:trudy_probability_positive}
\end{equation}
\end{subequations}
 
Therefore, a BNE can, in principle, be determined by considering candidate type-dependent support profiles and solving the corresponding systems of indifference equations in \eqref{eq:bob_support_conditions} and
\eqref{eq:trudy_support_conditions}, together with the probability constraints in \eqref{eq:bob_probability_constraints} and 
\eqref{eq:trudy_probability_constraints}. A candidate solution is an equilibrium only if all probabilities are feasible and the best-response inequalities for the actions outside the supports are satisfied.

\subsection{Precoder Codebook Design}
\label{subsec:codebDesignQ}
Since Trudy knows neither Alice's channel nor the RIS configuration selected by Bob, she cannot compute a configuration-specific precoder at the time of the attack. She therefore designs $\mathcal C_Q$ offline, based on the commonly known channel statistics and on Bob's codebook. For each training channel pair $(\bm F,\bm T)$ drawn from the joint channel distribution and each $\bm\omega\in\mathcal C_\Omega$, she computes the regularized least-squares precoder
\begin{equation}
\bm Q^\star
=
\left(\bm C_T^H\bm C_T+\lambda\bm I\right)^{-1}\bm C_T^H\bm C_A,
\label{eq:Qstar_train}
\end{equation}
with $\bm C_A=\bm G\bm\Omega\bm F$, $\bm C_T=\bm G\bm\Omega\bm T$, and $\lambda=10^{-3}\|\bm C_T\|_F^2$, which minimizes the mismatch between the two cascaded channels for that pair and configuration. The resulting set of precoders $\mathcal Q_T$ is then quantized into $N_Q-1$ codewords ($N_Q$ when the configuration-agnostic precoder is not included) through the Linde-Buzo-Gray (LBG) algorithm~\cite{lbg1980}, which minimizes the average distortion
\begin{equation}
\frac{1}{|\mathcal Q_T|}\sum_{\bm Q^\star\in\mathcal Q_T}
\min_{\bm Q\in\mathcal C_Q}\|\bm Q^\star-\bm Q\|_F^2 .
\end{equation}
Each codeword thus represents a cluster of channel and configuration realizations rather than a specific one, which captures the loss of adaptation induced by Trudy's partial knowledge.

\section{Numerical Results}
\label{sec:res}

In this section, we assess the performance of \ac{cr-pla} in the considered setting, under partial channel knowledge. We first describe the design of the finite action codebooks used by Bob and Trudy. We then evaluate the equilibrium MDP $V_{\mathrm{MD}}$ in~\eqref{eq:VMD} under the partial channel knowledge scenario. For each channel realization, the game is solved exactly by linear programming on the closed-form payoffs of Section~\ref{sec:PLA}, and the reported values of $V_{\rm MD}$ are averages over independent realizations; hence, values far below the inverse of the number of realizations remain meaningful and are not limited by the observation of misdetection events.

We compare the proposed mixed-strategy equilibrium against the following five benchmarks:
\begin{itemize}

\item \emph{Uniform Bob + BR Trudy}: Bob randomizes uniformly over the RIS codebook $\mathcal{C}_\Omega$, while Trudy plays her \ac{br} over $\mathcal{C}_Q$.

\item \emph{Pure Bob + BR Trudy}: Bob deterministically employs his best pure RIS configuration, while Trudy plays her \ac{br} over $\mathcal{C}_Q$.

\item \emph{Uniform Bob + Uniform Trudy}: both players randomize uniformly over their respective codebooks, with no best-response computation on either side.

\item \emph{BR Bob + Pure Trudy}: Trudy commits to a single precoder in $\mathcal{C}_Q$, while Bob plays his type-dependent \ac{br}; this is not a realistic policy, but provides a lower bound on $V_{\mathrm{MD}}$. 

\item \emph{Uniform Bob + Fixed Trudy}~\cite{tomasin2024analysis}: Bob randomizes uniformly over $\mathcal{C}_\Omega$, while Trudy employs, for each of her channel types, the configuration-agnostic precoder included in $\mathcal{C}_Q$, obtained by extending the fixed attack of~\cite{tomasin2024analysis} to the considered setting, as detailed in~\eqref{eq:Fhat_pure}--\eqref{eq:Qpure}.

\end{itemize}

To further investigate the robustness of the obtained results with respect to the adopted propagation model, we consider both the geometric multipath model of \cite{partI} and the  3GPP TR~38.901 urban microcell non-line-of-sight (UMi-NLOS) model \cite{3gpp38901, jaeckel2014quadriga} that takes into account effects of scattering.

\subsection{Codebook Design}
\label{subsec:codebDesignOmega}

\paragraph*{RIS codebook} As considered in \cite{partI}, Bob's codebook $\mathcal C_\Omega$ comprises $N_\Omega\le N$ columns of the unnormalized $N\times N$ DFT matrix, i.e., $[\bm\omega_k]_n=e^{-\jmath2\pi n c_k/N}$, $n=0,\ldots,N-1$, where the column indices $c_k$, $k=1,\ldots,N_\Omega$, are (approximately) equally spaced over $\{0,\ldots,N-1\}$. These unit-modulus vectors are mutually orthogonal and thus provide a structured set of distinct challenges: a precoder that reproduces Alice's cascaded channel for one configuration is generally ineffective for the others, which Bob exploits through randomization. Since this design does not account for the legitimate rate, in the following we also consider codebooks obtained by perturbing the communication-optimal RIS configuration.

\subsection{Channel Model}
\label{sec:VB}

To take into account the information about $\bm F$ that Trudy can indirectly obtain through her own channel, we model $\bm T$ as
\begin{equation}
\label{eq:correlated_T}
\begin{split}
    \bm T(:,j)
    ={}&
    \rho\,
    \bm F
    \left(
    :,
    ((j-1)\bmod K)+1
    \right)                                                         \\
    &+
    \sqrt{1-\rho^2}\,
    \overline{\bm T}(:,j),
    \qquad
    j=1,\ldots,N_{\rm T}\,,
\end{split}
\end{equation}
where $\rho\in[0,1]$ controls the dependence between the two channels. Since $\bm F$ and $\overline{\bm T}$ are independent, zero-mean, and have equal per-entry power, $\rho$ coincides with the correlation coefficient between corresponding entries of $\bm F$ and $\bm T$, ranging from independent channels ($\rho=0$) to maximal dependence ($\rho=1$). Although this model does not account for any geometric relation between the Alice-\ac{RIS} and the Trudy-\ac{RIS} channels, it has already been considered, for example in~\cite{tomasin2024analysis}. For $K=1$, all columns of $\bm T$ share the same correlated component, so that $\operatorname{rank}(\bm T)\le 1+\operatorname{rank}(\overline{\bm T})$. By observing $\bm T$, Trudy thus obtains indirect information on $\bm F$ through $p(\bm F\mid\bm T)$, whereas Bob's uncertainty on Trudy's channel is described by $p(\bm T\mid\bm F)$.

Note that~\eqref{eq:correlated_T} is a statistical, rather than geometric, model of the channel dependence. For $\rho=1$, the independent component vanishes and, for $K=1$, all columns of $\bm T$ coincide, so that $\operatorname{rank}(\bm T)=1$ irrespective of $N_{\rm T}$. Hence, $\rho=1$ is the maximally correlated limit of the model: each player can infer the opponent's channel from its own, so that the game reduces to one with complete information, which differs from the perfect-knowledge setting of Part~I only by the rank-one structure imposed on $\bm T$.

For the geometric model, the type sets contain $N_{\mathcal F}=N_{\mathcal T}=4$ realizations, and each $\bm T_\ell$ is generated through~\eqref{eq:correlated_T} from $\bm F_\ell$ and an independent realization $\overline{\bm T}_\ell$. Approximating the entries of $\overline{\bm T}$ as i.i.d. zero-mean circularly-symmetric Gaussian with unit power, \eqref{eq:correlated_T} yields, for $0<\rho<1$ and $K=1$,
\begin{equation}
\Pr(\bm T_\ell\mid\bm F_k)=\frac{\exp\!\left(-\|\bm T_\ell-\rho\bm F_k\bm 1^{\rm T}\|_F^2/(1-\rho^2)\right)}{\sum_{r=1}^{N_{\mathcal T}}\exp\!\left(-\|\bm T_r-\rho\bm F_k\bm 1^{\rm T}\|_F^2/(1-\rho^2)\right)},
\label{eq:prior_geom}
\end{equation}
while $\gamma_{k\ell}=1/(N_{\mathcal F}N_{\mathcal T})$ for $\rho=0$ and $\gamma_{k\ell}=\delta_{k\ell}/N_{\mathcal F}$ for $\rho=1$. The prior thus ranges from a factorized one, where each player's own type carries no information on the opponent's, to a diagonal one, where it reveals it exactly. Trudy's LBG codebook is trained on $20$ type pairs drawn according to $\gamma_{k\ell}$, each combined with all configurations in $\mathcal C_\Omega$.

For the 3GPP model, the type sets contain $N_{\mathcal F}=N_{\mathcal T}=2$ realizations. Alice's types are independent realizations of the propagation environment for a fixed Alice position, whereas Trudy's types are obtained at independent random positions within the considered distance interval, as detailed in Section~\ref{sec:3gpp_partial}. Since each link is generated by an independent QuaDRiGa realization, $\bm F$ and $\bm T$ are statistically independent, and the prior is exactly $\gamma_{k\ell}=1/(N_{\mathcal F}N_{\mathcal T})$; the relative position of Alice and Trudy affects the game only through the spatial structure of the generated channels. Trudy's LBG codebook is trained on all pairs in $\mathcal F\times\mathcal T$ and on all configurations in $\mathcal C_\Omega$.
In both cases, the game and its equilibrium value are evaluated on the finite type sets, so that channel realizations outside them are not considered. The type sets are redrawn at each independent realization, and each point is averaged over $300$ realizations for the geometric model and $100$ for the 3GPP model.

\subsection{Performance on Geometric Channel}

\begin{figure}
    \centering
    \includegraphics[width=\columnwidth]{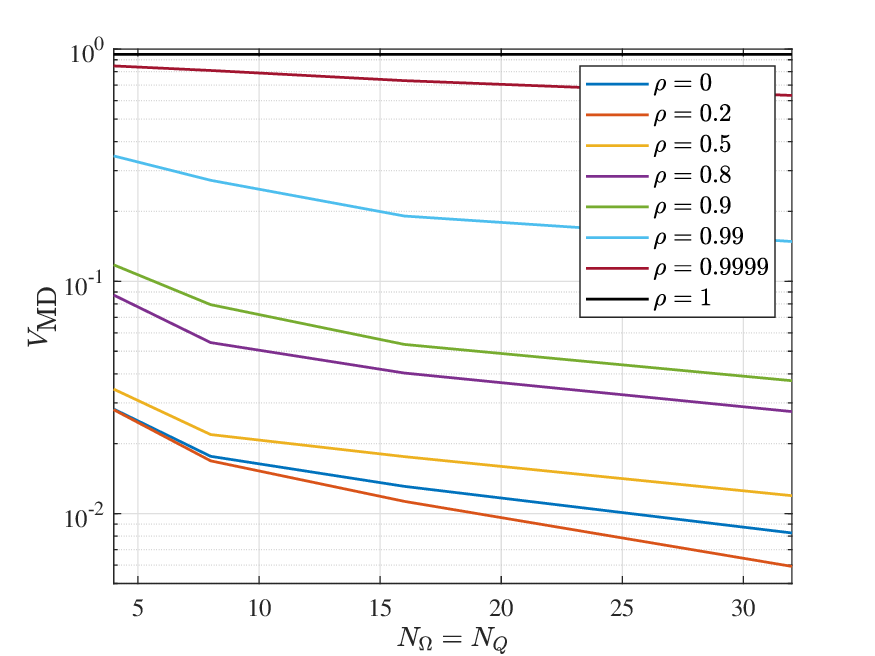}
    \caption{Misdetection probability $V_{\mathrm{MD}}$  as a function of Bob's and Trudy's codebook dimension, with $N_\Omega=N_Q$ and $L_G= 1$. $M=4$, SNR$=10$ dB, $N_{\rm T} =1$, $L_T = L_F =3$. Target $P_{\rm FA} = 5\cdot10^{-2}$.}
    \label{fig:pmd_vs_rho}
\end{figure}
\begin{figure}
    \centering
    \includegraphics[width=\columnwidth]{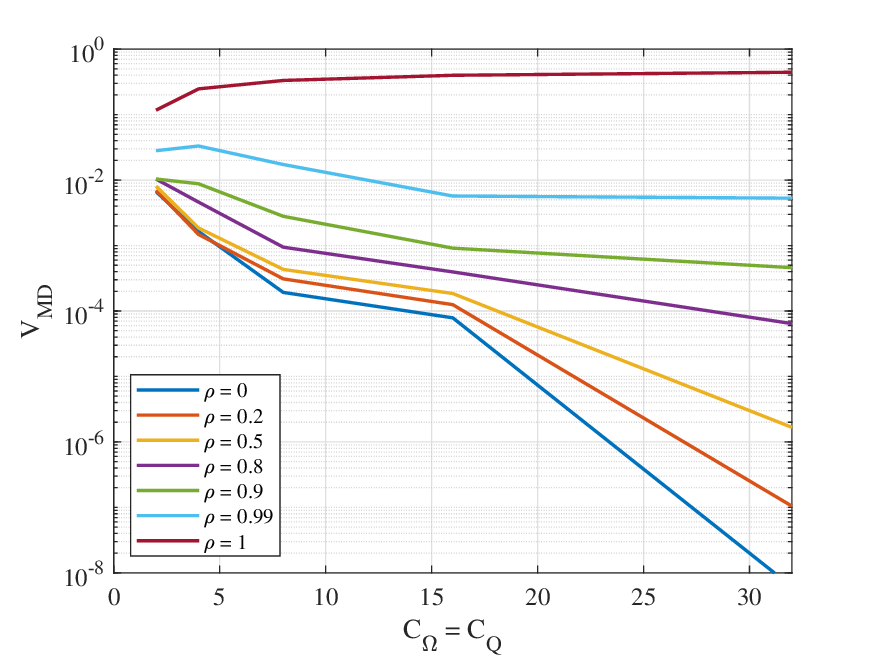}
    \caption{Misdetection probability $V_{\mathrm{MD}}$  as a function of Bob's and Trudy's codebook dimension, with $N_\Omega=N_Q$, and $L_G= 3$. $M=4$, SNR$=10$ dB, $N_{\rm T} =1$, $L_T = L_F =3$. Target $P_{\rm FA} = 5\cdot10^{-2}$.}
    \label{fig:pmd_vs_rho_LG3}
\end{figure}
\begin{figure}
    \centering
    \includegraphics[width=\columnwidth]{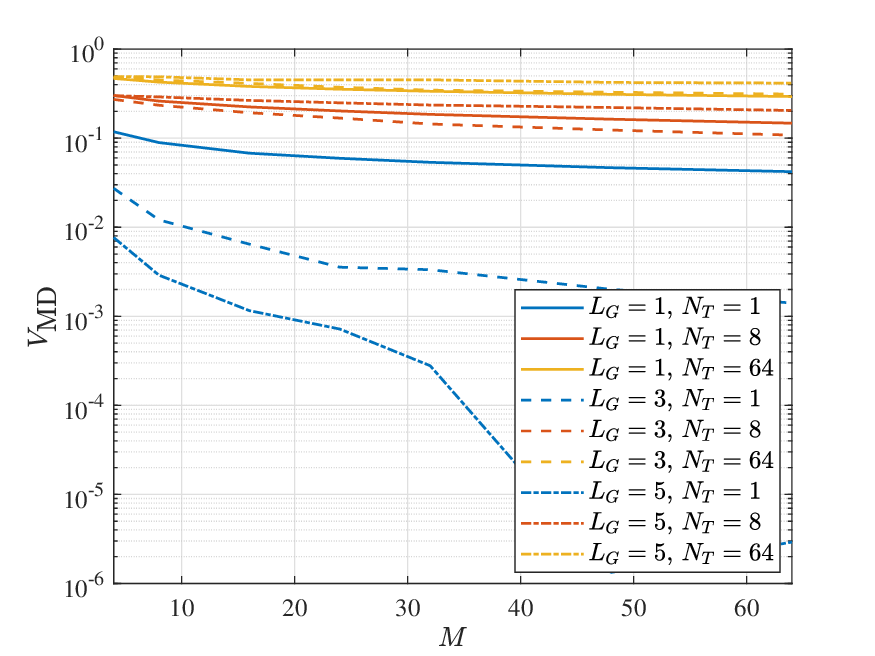}
    \caption{Misdetection probability $V_{\mathrm{MD}}$ as a function of the number of antennas at Bob $M$, with $N_\Omega = N_Q = 32$. $N=64$, $\rho = 0.5$, SNR$=-5$ dB, $L_T = L_F =3$. Target $P_{\rm FA} = 5\cdot10^{-2}$.}
    \label{fig:pmd_vs_M}
\end{figure}
\begin{figure}
    \centering
    \includegraphics[width=\columnwidth]{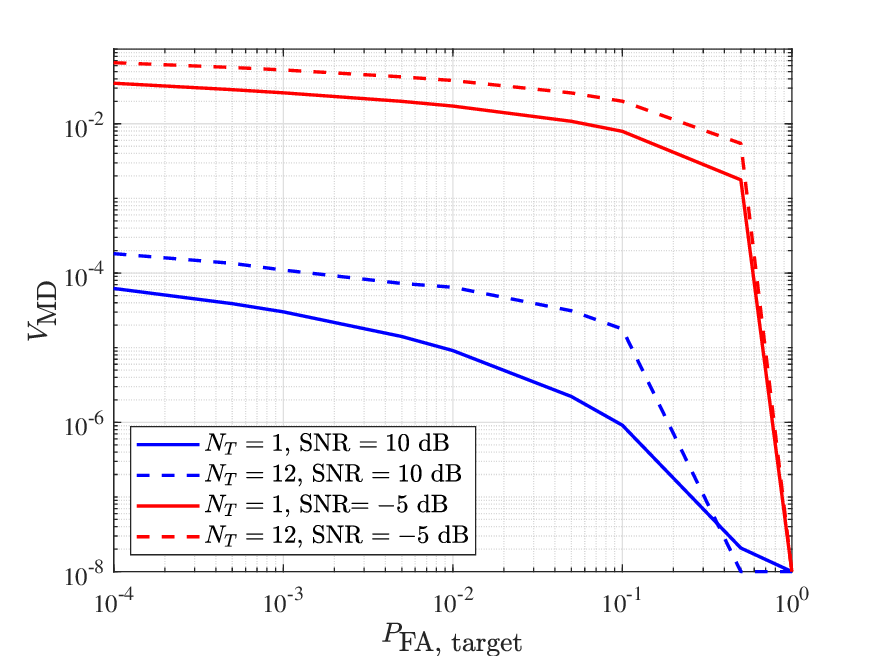}
    \caption{Misdetection probability $V_{\mathrm{MD}}$ as a function of the false alarm probability $P_{\mathrm{FA}}$, with $N =32$ RIS elements, $N_\Omega = N_Q = 16$. $M=8$, $\rho = 0$, $L_G = 3$, $L_T = L_F =3$.}
    \label{fig:pmd_vs_pfa}
\end{figure}

We consider here the geometric channel model of \cite{partI}, where the transmitted signal reaches the receiver through a small number of paths (mmWave scenario). Let $L_F$ ($L_G$) denote the number of paths for the Alice-RIS (RIS-Bob) channel and $L_T$ the number of paths for $\overlinebold{T}$. We suppose that the RIS has $N=64$ reflecting elements, unless otherwise stated, while Alice is equipped with a single antenna ($K=1$). 
Figs.~\ref{fig:pmd_vs_rho} and~\ref{fig:pmd_vs_rho_LG3} show $V_{\rm MD}$ as a function of the codebook size $N_\Omega=N_Q$, increased jointly up to $32$, for different values of $\rho$ and for $L_G=1$ and $L_G=3$, respectively, with $N_{\rm T}=1$, and $M=4$. The configuration-agnostic precoder is not included in Trudy's codebook in Figs.~\ref{fig:pmd_vs_rho}--\ref{fig:pmd_vs_pfa}, since \eqref{eq:Fhat_pure} is undefined for $\rho=0$ and, for $\rho>0$, we verified that its inclusion has no impact on $V_{\rm MD}$.
As expected, $V_{\rm MD}$ increases with $\rho$, since Trudy's channel becomes statistically closer to Alice's. For $\rho=1$, $\bm T=\bm F$, so that all training precoders coincide, up to the regularization, with $\bm Q=1$, which yields a zero mismatch for every RIS configuration; hence, $V_{\rm MD}$ reaches $1-P_{\rm FA}$ regardless of the codebook size. For $\rho<1$, larger codebooks favor Bob and reduce $V_{\rm MD}$. The number of RIS--Bob paths also plays an important role: with $L_G=1$, $V_{\rm MD}$ never falls below $5\cdot10^{-2}$, whereas with $L_G=3$ it reaches much lower values (less than $10^{-10}$) for uncorrelated channels.

In Fig. \ref{fig:pmd_vs_M}, we show the misdetection probability $V_{\textrm {MD}}$ as a function of the number of receiving antennas at Bob, $M$, for different combinations of the number of paths in the RIS--Bob channel, $L_G \in \{1,3,5\}$, and of the number of antennas at Trudy, $N_{\rm T} \in \{1,8,64\}$. Both codebook dimensions are fixed to $N_\Omega = N_Q = 32$ and the correlation coefficient is set to $\rho=0.5$. The dominant parameter is clearly 
$N_{\rm T}$. When Trudy is equipped with multiple antennas, the curves collapse into a narrow band, becoming virtually flat with respect to $M$ and nearly indistinguishable across different values of $L_G$, especially with $N_{\rm T}=64$. Only for $N_{\rm T} = 1$, the remaining parameters retain a visible influence, and in that case both a richer RIS--Bob scattering and a larger number of receiving antennas translate into a substantial reduction of the misdetection probability: with $L_G =5$, $V_{\textrm {MD}}$ decreases from approximately $10^{-2}$ at $M=4$ to about $10^{-6}$ at $M=64$, whereas with $L_G=1$ the same increase in $M$ does not have the same impact. The saturation observed for $N_{\rm T}>1$ indicates that little deterministic mismatch is left for Bob to resolve: the diversity offered by the RIS--Bob channel and by Bob's array can only be exploited if a residual difference between the two cascaded channels exists in the first place.
Increasing $M$ is an effective countermeasure only as long as Trudy's spatial resources remain limited.

Fig.~\ref{fig:pmd_vs_pfa} shows the misdetection probability $V_{\mathrm{MD}}$ as a function of the target false alarm probability $P_{\mathrm{FA}}$, for two values of $N_{\rm T}$ and two SNR values, with $N=32$ RIS elements, $N_\Omega = N_Q = 16$, $M=8$ and uncorrelated channels ($\rho=0$). The curves describe the \ac{det} of the authentication test: relaxing the false alarm constraint monotonically reduces $V_{\mathrm{MD}}$, with a slow decay up to $P_{\mathrm{FA}} \approx 10^{-1}$ followed by a more evident drop as $P_{\mathrm{FA}}$ approaches one, where all curves tend to zero. The SNR is the dominant parameter over the whole range, separating the two groups of curves by about three orders of magnitude, while the effect of $N_{\rm T}$ is comparatively modest and vanishes in the steep region: since the channels are uncorrelated, additional antennas at Trudy provide no correlated component to be exploited and only marginally improve the attack. The knee of the curves shifts to higher $P_{\mathrm{FA}}$ as the SNR decreases, indicating that at low SNR the false alarm constraint must be considerably relaxed before any appreciable gain in detection is obtained.

\begin{figure}[]
    \centering
    \includegraphics[width=0.45\textwidth]{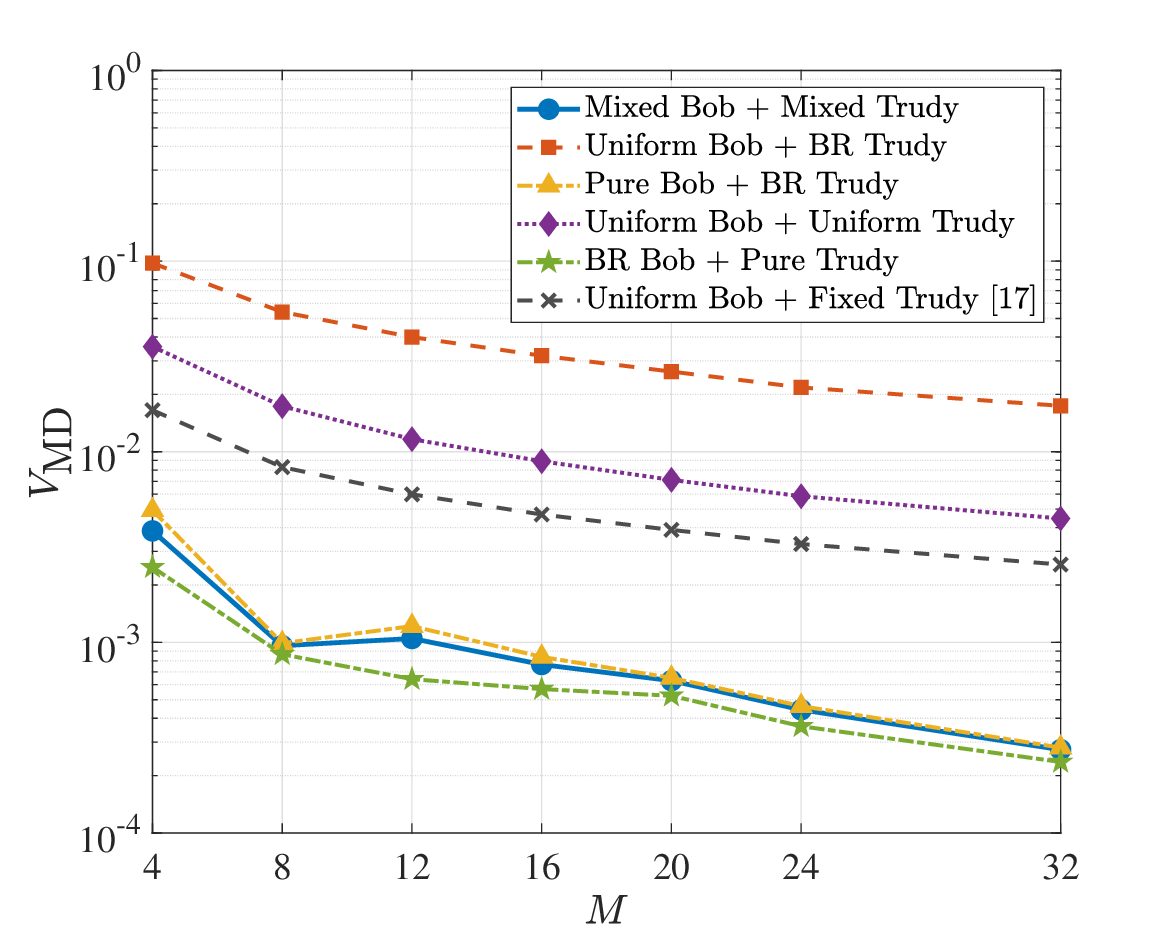}
    \caption{Comparison between the proposed BNE strategy (Mixed Bob + Mixed Trudy) and the considered baseline policies as a function of $M$, for $N_{\rm T}=8$, $L_T=7$, $L_G=6$, $N = 32$, $N_\Omega=4$, $N_Q=5$, $\mathrm{SNR}=5$~dB, $P_{\mathrm{FA}}=10^{-2}$, and $\rho = 0.1$.
    }
    \label{fig:benchmark_bay_rho01_Lg6}
\end{figure}

\begin{figure}[]
    \centering
    \includegraphics[width=0.45\textwidth]{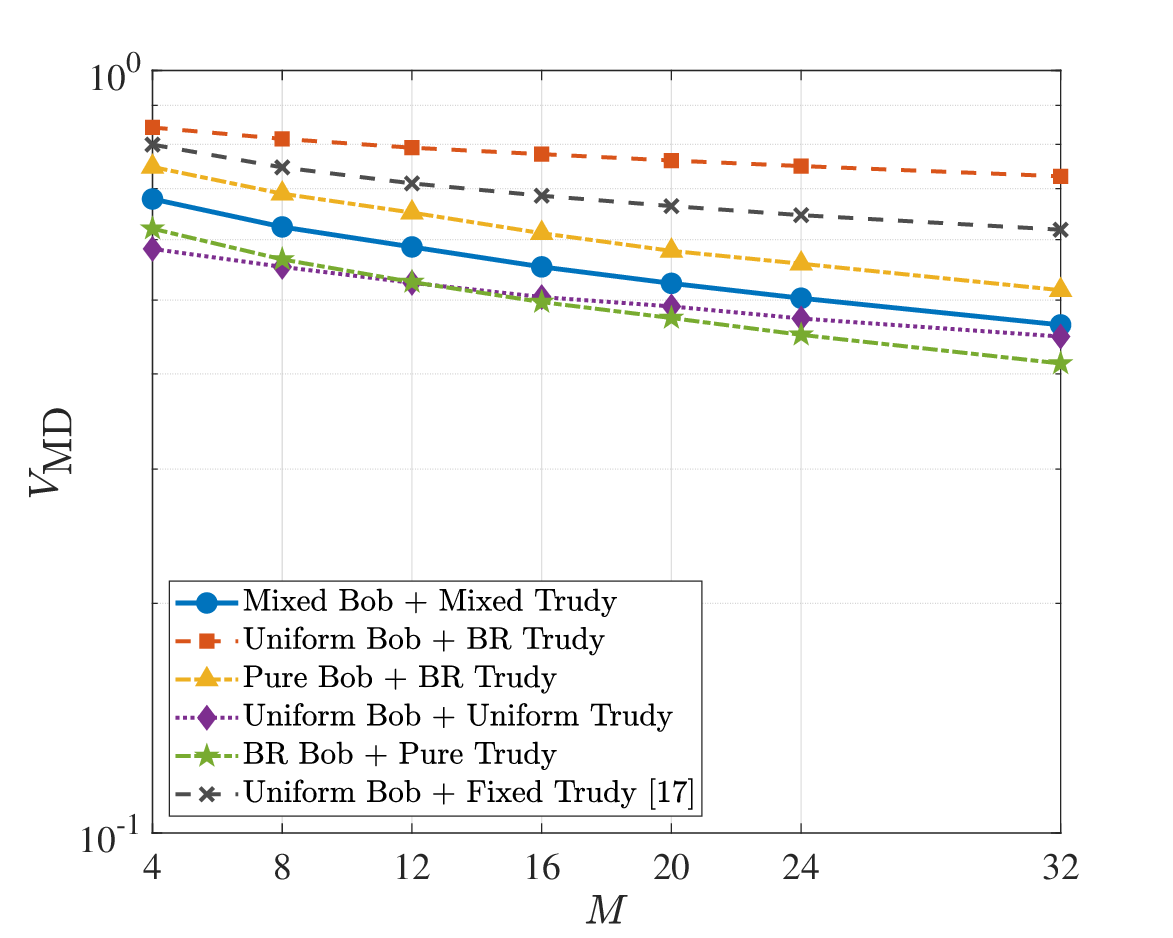}
    \caption{Comparison between the proposed BNE strategy (Mixed Bob + Mixed Trudy) and the considered baseline policies as a function of $M$, for $N_{\rm T}=8$, $L_T=7$, $L_G=6$, $N = 32$, $N_\Omega=4$, $N_Q=5$, $\mathrm{SNR}=5$~dB, $P_{\mathrm{FA}}=10^{-2}$, and $\rho = 0.9$.
    }
    \label{fig:benchmark_bay_rho09_Lg6}
\end{figure}

\begin{figure}[]
    \centering
    \includegraphics[width=0.45\textwidth]{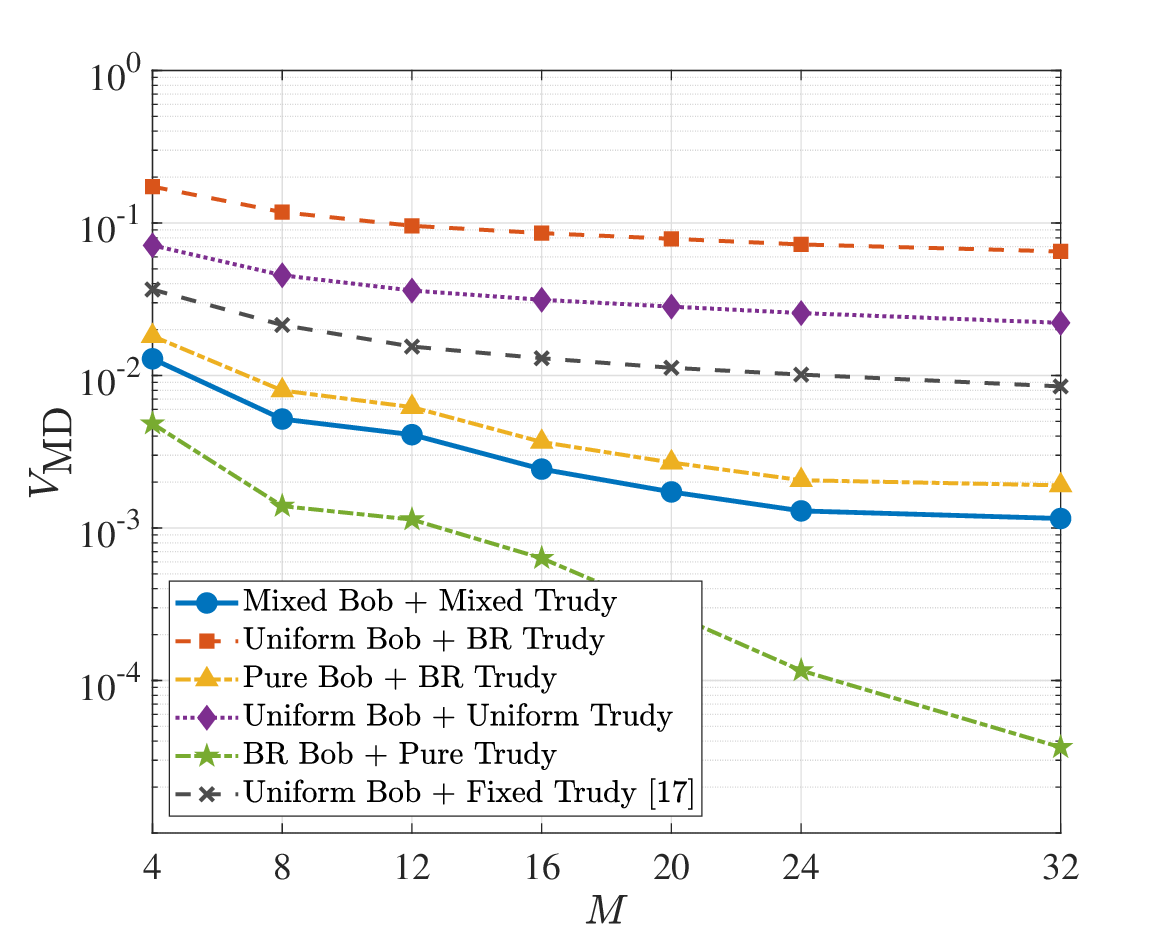}
    \caption{Comparison between the proposed BNE strategy (Mixed Bob + Mixed Trudy) and the considered baseline policies as a function of $M$, for $N_{\rm T}=8$, $L_T=7$, $L_G=3$, $N = 32$, $N_\Omega=4$, $N_Q=5$, $\mathrm{SNR}=5$~dB, $P_{\mathrm{FA}}=10^{-2}$, and $\rho = 0.1$.
    }
    \label{fig:benchmark_bay_rho01}
\end{figure}

\begin{figure}[]
    \centering
    \includegraphics[width=0.45\textwidth]{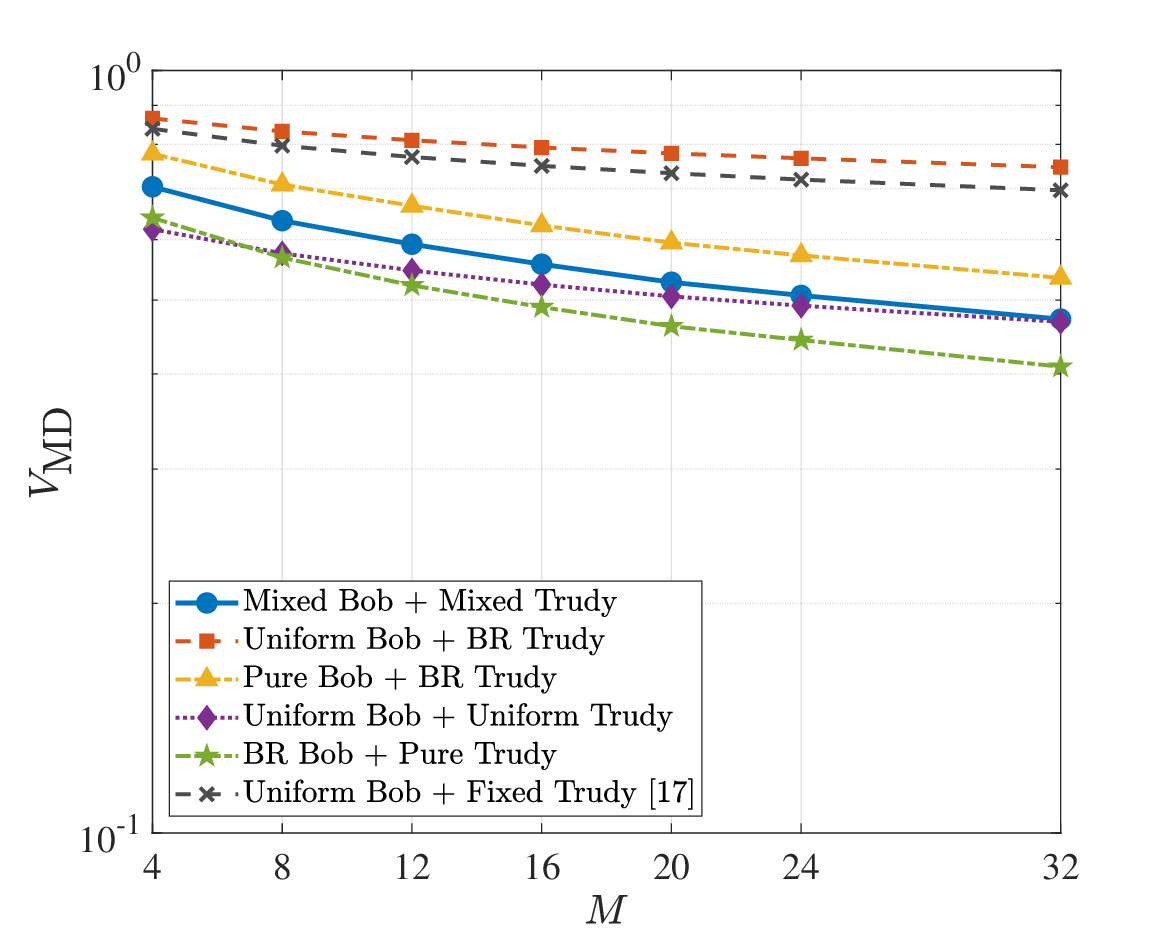}
    \caption{Comparison between the proposed BNE strategy (Mixed Bob + Mixed Trudy) and the considered baseline policies as a function of $M$, for $N_{\rm T}=8$, $L_T=7$, $L_G=3$, $N = 32$, $N_\Omega = 4$, $N_Q=5$, $\mathrm{SNR}=5$~dB, $P_{\mathrm{FA}}=10^{-2}$, and $\rho = 0.9$.
    }
    \label{fig:benchmark_bay_rho09}
\end{figure}

Figs.~\ref{fig:benchmark_bay_rho01_Lg6}, \ref{fig:benchmark_bay_rho09_Lg6}, \ref{fig:benchmark_bay_rho01}, and~\ref{fig:benchmark_bay_rho09} compare the BNE (Mixed Bob + Mixed Trudy) with the baselines introduced at the beginning of this section, for $N_{\rm T}=8$, $N = 32$, $L_T=7$, $N_\Omega=N_Q-1 = 4$, $\mathrm{SNR}=5$~dB, $P_{\mathrm{FA}}=10^{-2}$, $L_G\in\{3,6\}$, and $\rho\in\{0.1,0.9\}$. Although $\min(N_{\rm T},L_T)=7$ exceeds both values of $L_G$, which under perfect knowledge would allow Trudy to forge any deterministic RIS configuration (see \cite{partI}), the pure-Bob baseline, here a type-dependent choice of the RIS configuration, does not reach $1-P_{\mathrm{FA}}$, since Trudy's precoders are designed offline without knowledge of Bob's configuration or type.

In the \emph{Uniform Bob + Fixed Trudy} baseline, Bob randomizes uniformly over $\mathcal{C}_\Omega$, while Trudy employs a single, configuration-agnostic precoder for each of her channel types. 
For each type $\bm T_\ell$, Trudy first estimates Alice's channel by averaging her $N_{\rm T}$ correlated observations, i.e., the columns of $\bm T_\ell$ in~\eqref{eq:correlated_T},
\begin{equation}
\hat{\bm F}_\ell = \frac{1}{\rho N_{\rm T}} \sum_{j=1}^{N_{\rm T}} \left[\bm T_\ell\right]_{(:,j)},
\label{eq:Fhat_pure}
\end{equation}
and then reproduces this estimate through her own channel by solving $\bm T_\ell \bm Q = \hat{\bm F}_\ell$ in the least-squares, minimum-norm sense,
\begin{equation}
\bm Q_{\mathrm{agn},\ell} = \bm T_\ell^{\dagger} \hat{\bm F}_\ell.
\label{eq:Qpure}
\end{equation}
This construction generalizes the fixed attack of~\cite[Eq.~(35)]{tomasin2024analysis}, derived for a single-antenna legitimate user, a single-antenna verifier, and an attacker with as many antennas as RIS elements, to the multiple-antenna setting considered here. When $N_{\rm T}=N$, $\bm T_\ell$ is generically full row rank, $\bm T_\ell\bm Q_{\mathrm{agn},\ell}=\hat{\bm F}_\ell$ for every RIS configuration, and the only residual mismatch is due to the estimation error $\hat{\bm F}_\ell-\bm F$, which vanishes as $\rho\to1$. For $N_{\rm T}<N$, a further residual mismatch arises from the rank deficiency of $\bm T_\ell$. 

All the baselines lie within the action space $\mathcal C_\Omega\times\mathcal C_Q$ of the game, since $\mathcal C_Q$ includes, for each of Trudy’s types, the configuration-agnostic precoder \eqref{eq:Qpure}.
Since the \ac{br} to a fixed opponent strategy is attained by a pure strategy, Pure Bob + BR Trudy and Uniform Bob + BR Trudy yield an upper bound to $V_{\mathrm{MD}}$, whereas BR Bob + Pure Trudy yields a lower bound to $V_{\mathrm{MD}}$. 
Conversely, Uniform Bob + Uniform Trudy and Uniform Bob + Fixed Trudy carry no guaranteed ordering with respect to $V_{\mathrm{MD}}$, since neither involves a BR, although the latter is upper bounded by Uniform Bob + BR Trudy.
Indeed, the former falls below the equilibrium value for $\rho=0.9$ over most of the range of $M$, while the latter remains close to or above it in all cases, suggesting that the loss incurred by Bob with uniform randomization outweighs Trudy's advantage from the configuration-agnostic precoder. 
$V_{\textrm{MD}}$ is lower for $L_G=6$ than for $L_G=3$, by a factor of about $3$--$4$ for $\rho=0.1$ (Figs.~\ref{fig:benchmark_bay_rho01_Lg6} and~\ref{fig:benchmark_bay_rho01}) and only slightly for $\rho=0.9$ (Figs.~\ref{fig:benchmark_bay_rho09_Lg6} and~\ref{fig:benchmark_bay_rho09}), confirming that the number of RIS--Bob paths is an important factor under partial channel knowledge and quantized attacker precoders, especially with low correlation.  The effect of correlation is also shown by comparing $\rho = 0.1$ with $\rho = 0.9$. Increasing $\rho$ raises $V_{\textrm{MD}}$ across all baselines for both values of $L_G$, since a higher correlation makes Trudy's channel statistically closer to Alice's, reducing the mismatch that Bob can exploit for discrimination even with a quantized attacker precoder. This effect is even stronger for $L_G=6$, where $V_{\textrm{MD}}$ increases by about three orders of magnitude at $M=32$, against about $300$ times for $L_G=3$. For $\rho=0.9$, the game value remains close to the Uniform Bob + Uniform Trudy baseline for both values of $L_G$, indicating that under strong correlation randomizing the RIS configuration yields only a marginal security improvement over a naive uniform policy.

\paragraph*{Effect of the RIS codebook design}

\begin{figure}[]
    \centering
    \includegraphics[width=\columnwidth]{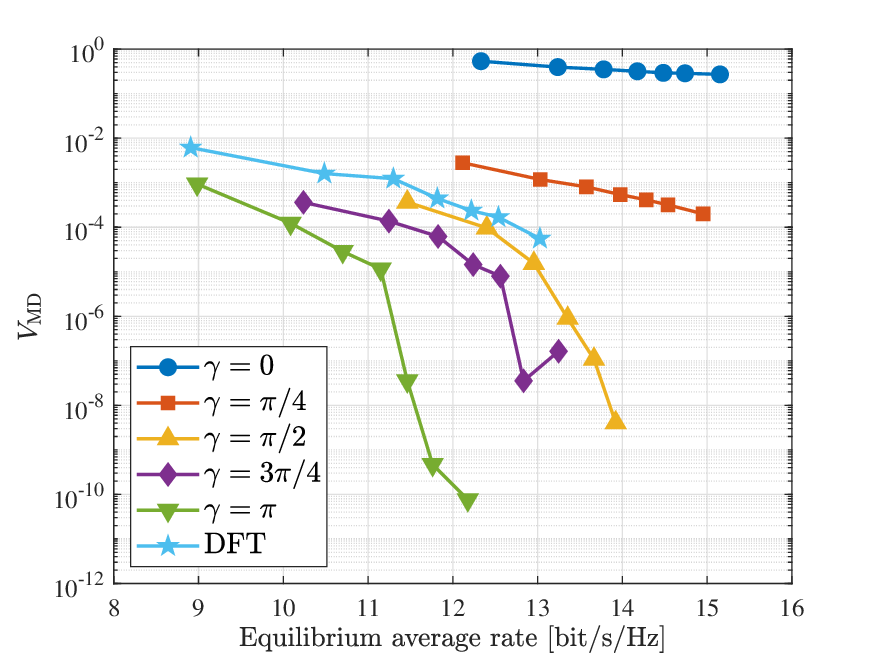}
    \caption{$V_{\rm MD}$ as a function of the equilibrium-weighted average legitimate rate for the DFT-based RIS codebook and codebooks uniformly perturbed around the communication-optimal configuration, $N_{\rm T}=12$, $L_T=3$, $L_G=3$, $N=32$, $N_\Omega =N_Q = 16$, $\mathrm{SNR}=5$~dB, $\rho=0$, and $P_{\rm FA}=10^{-2}$. The points along each curve correspond to $M\in\{4,8,12,16,20,24,32\}$.
    }
    \label{fig:phiOptComparison_bayesian}
\end{figure}

The DFT-based codebook of Section~\ref{subsec:codebDesignOmega} is designed for authentication only and does not account for the legitimate communication rate. To assess the resulting rate--security tradeoff, we also consider RIS codebooks generated around the communication-optimal configuration $\overline{\bm\omega}(\bm F)$, i.e., the RIS configuration that maximizes the legitimate rate for the Alice--RIS channel realization $\bm F$. Following~\cite{tomasin2024analysis}, the $k$-th codeword is obtained as
\begin{equation}
\bm\omega_k(\gamma)
=
\overline{\bm\omega}(\bm F)
\odot
\exp\left(\jmath\bm\delta_k\right),
\qquad
[\bm\delta_k]_n\sim\mathcal U[-\gamma,\gamma],
\label{eq:perturbed_codebook}
\end{equation}
where $\odot$ denotes the Hadamard product, the phase perturbations are independently generated across RIS elements and codewords, and $\gamma\in[0,\pi]$ controls the departure from $\overline{\bm\omega}(\bm F)$. For $\gamma=0$, all codewords coincide with the communication-optimal configuration, which maximizes the rate but provides no challenge diversity; for $\gamma=\pi$, all phases are uniformly distributed over the unit circle, and the codebook retains no statistical preference for $\overline{\bm\omega}(\bm F)$. Since $\overline{\bm\omega}(\bm F)$ depends on Alice's channel, \eqref{eq:perturbed_codebook} is applied independently for each type $\bm F\in\mathcal F$, yielding a type-dependent RIS codebook, whereas Trudy's codebook $\mathcal C_Q$ is trained offline through the LBG algorithm of Section~\ref{subsec:codebDesignQ} by pooling the training samples across all types.

Fig.~\ref{fig:phiOptComparison_bayesian} shows the tradeoff between the equilibrium-weighted average legitimate rate and $V_{\rm MD}$ for the DFT-based codebook and for the perturbed codebooks with $\gamma\in\{0,\pi/4,\pi/2,3\pi/4,\pi\}$, with $L_G=3$, $L_T=3$, $N_{\rm T}=12$, $N=32$, $N_\Omega = N_Q = 16$, $\mathrm{SNR}=5$~dB, $\rho = 0$, and $P_{\rm FA}=10^{-2}$. In this figure, we do not consider the presence of the agnostic precoder in Trudy's codebook. The legitimate rate of each RIS configuration is the water-filling capacity of the cascaded channel $\bm G\bm\Omega\bm F$; among the equilibrium strategies of Bob attaining $V_{\rm MD}$, we select the one maximizing the type-averaged rate. The points along each curve correspond to $M\in\{4,8,12,16,20,24,32\}$. 

For $\gamma=0$, Bob provides no configuration diversity, and $V_{\rm MD}$ remains close to the indistinguishability limit $1-P_{\rm FA}$, despite this design yielding the largest rates. Already a moderate perturbation ($\gamma=\pi/4$) produces a substantial security gain, and increasing $\gamma$ further progressively trades communication rate for security. The codebook with $\gamma=\pi$ achieves the lowest $V_{\rm MD}$, down to about $10^{-10}$ at $M=32$, at the cost of the lowest rates among the perturbed codebooks. For all values of $\gamma$, $V_{\rm MD}$ decreases with $M$, except for $\gamma=3\pi/4$, which shows an increase at the largest value of $M$.

The codebook with $\gamma=\pi$ also achieves a markedly lower $V_{\rm MD}$ than the DFT-based codebook over the whole considered range of $M$, with a gap that widens as $M$ increases and reaches several orders of magnitude for the largest arrays. A similar, although less pronounced, behavior is observed for $\gamma=\pi/2$ and $\gamma=3\pi/4$. The mutual orthogonality of the DFT configurations is mainly beneficial against an attacker that tailors her precoder to each individual configuration, since such a precoder is then poorly matched to the others. Here, instead, Trudy's precoders are trained offline and pooled across all types and RIS configurations, so that they are only loosely tied to the geometry of any single configuration, and the orthogonality of the DFT codebook provides comparatively little benefit.

\subsection{Performance on the 3GPP Channel}
\label{sec:3gpp_partial}

Finally, we assess the proposed framework in a standardized propagation environment, in which the Alice--RIS, Trudy--RIS, and RIS--Bob channels are generated according to the 3GPP TR~38.901 UMi-NLOS model~\cite{3gpp38901, jaeckel2014quadriga}. The carrier frequency is $28$~GHz, the RIS comprises $N=32$ reflecting elements, and Alice is equipped with a single antenna, i.e., $K=1$. We consider $M\in\{4,12,20,32\}$ antennas at Bob, $N_{\rm T}\in\{2, 6, \dots, 32\}$ antennas at Trudy, $N_\Omega = 8$, $N_Q=9$, target false-alarm probabilities $P_{\mathrm{FA}}=10^{-1}$ and $P_{\mathrm{FA}}=10^{-3}$, and SNR values in $\{-10,-5,0\}$~dB. We also considered $\mathrm{SNR}=5$ and $10$~dB, without reporting the results here, since $V_{\mathrm{MD}}$ falls below $10^{-10}$, the lowest value shown in the figures, for all the considered parameter combinations.

Unlike the geometric model, the correlation between the Alice--RIS and Trudy--RIS channels is not imposed through the coefficient $\rho$ in~\eqref{eq:correlated_T}, but results from the relative position of the two terminals. 

Trudy's position is randomly selected around Alice, with a uniformly distributed angular position and a distance falling within one of two intervals, namely 1--2~m and 10--12~m. 
Each link is generated by an independent QuaDRiGa realization, and their dependence is purely geometric: 
the first interval yields strongly correlated channels, whereas the second yields weakly correlated ones. Since the two intervals entail similar path losses towards the RIS, their comparison mainly captures differences in the spatial and multipath structure of the channel realizations rather than in the received power.
In these simulations, we do not exploit the spatial consistency of QuaDRiGa, which would additionally correlate the large-scale and cluster-level parameters of nearby terminals. We verified that generating the Alice–RIS and Trudy–RIS channels within a spatially consistent realization, with Trudy at 1–2 m from Alice, increases $V_{\rm MD}$ by a factor of about 2–3, while leaving the qualitative behavior unchanged and $V_{\rm MD}$ below $0.1$ over the whole considered range.

\begin{figure}[]
    \centering
    \includegraphics[width=0.50\textwidth]{{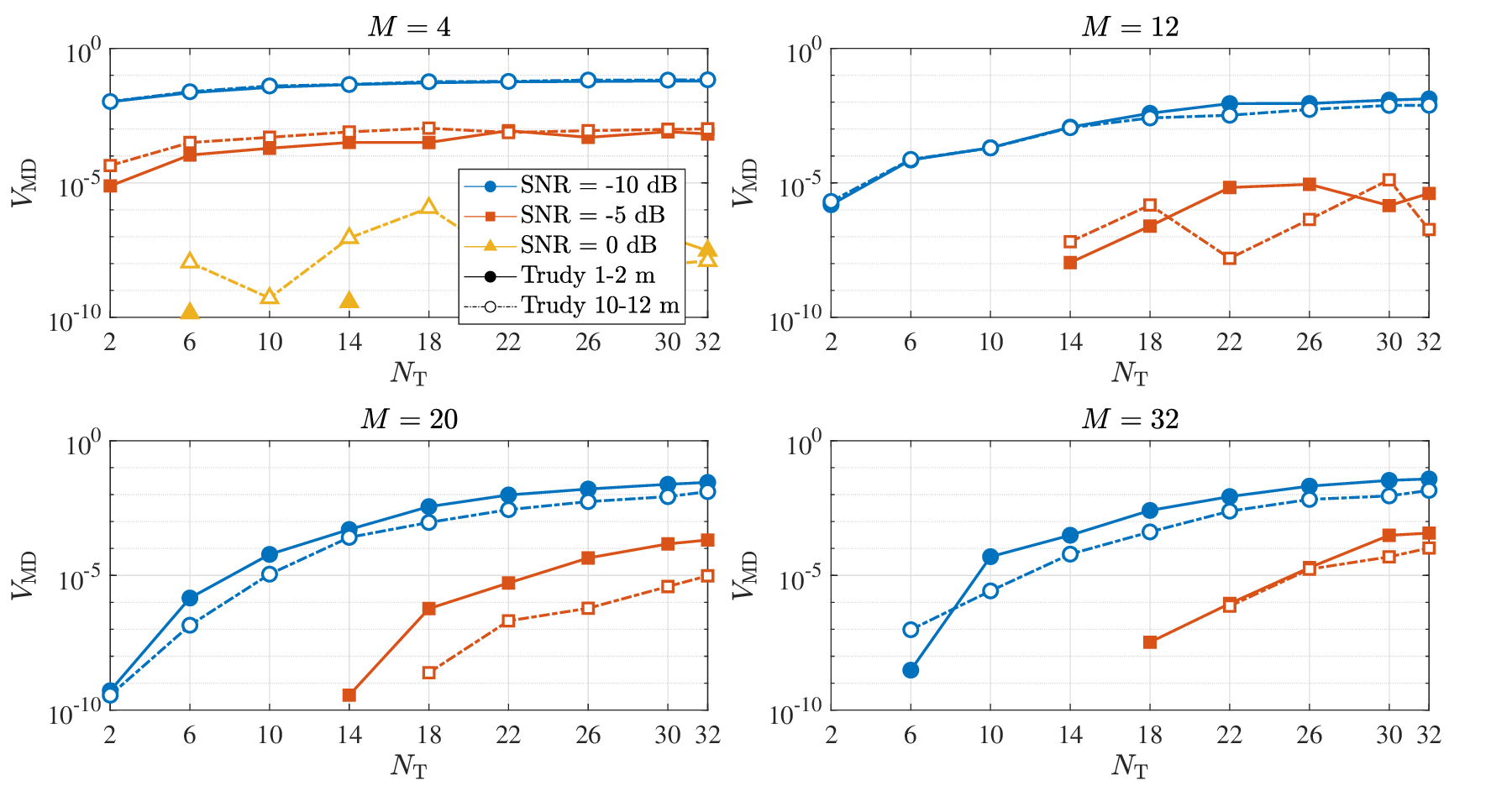}}
    \caption{
    $V_{\mathrm{MD}}$ as a function of the number of antennas at Trudy, $N_{\rm T}$, for $P_{\mathrm{FA}}=10^{-1}$.
    The four panels correspond to different numbers of receive antennas at Bob, $M\in\{4,12,20,32\}$.
    Colors and markers identify the considered SNR values, while filled and empty markers refer to the Trudy--Alice distance ranges $1$--$2$ m and  $10$--$12$ m, respectively.
    }
    \label{fig:pmd_pfa_1}
\end{figure}

\begin{figure}[]
    \centering
    \includegraphics[width=0.50\textwidth]{{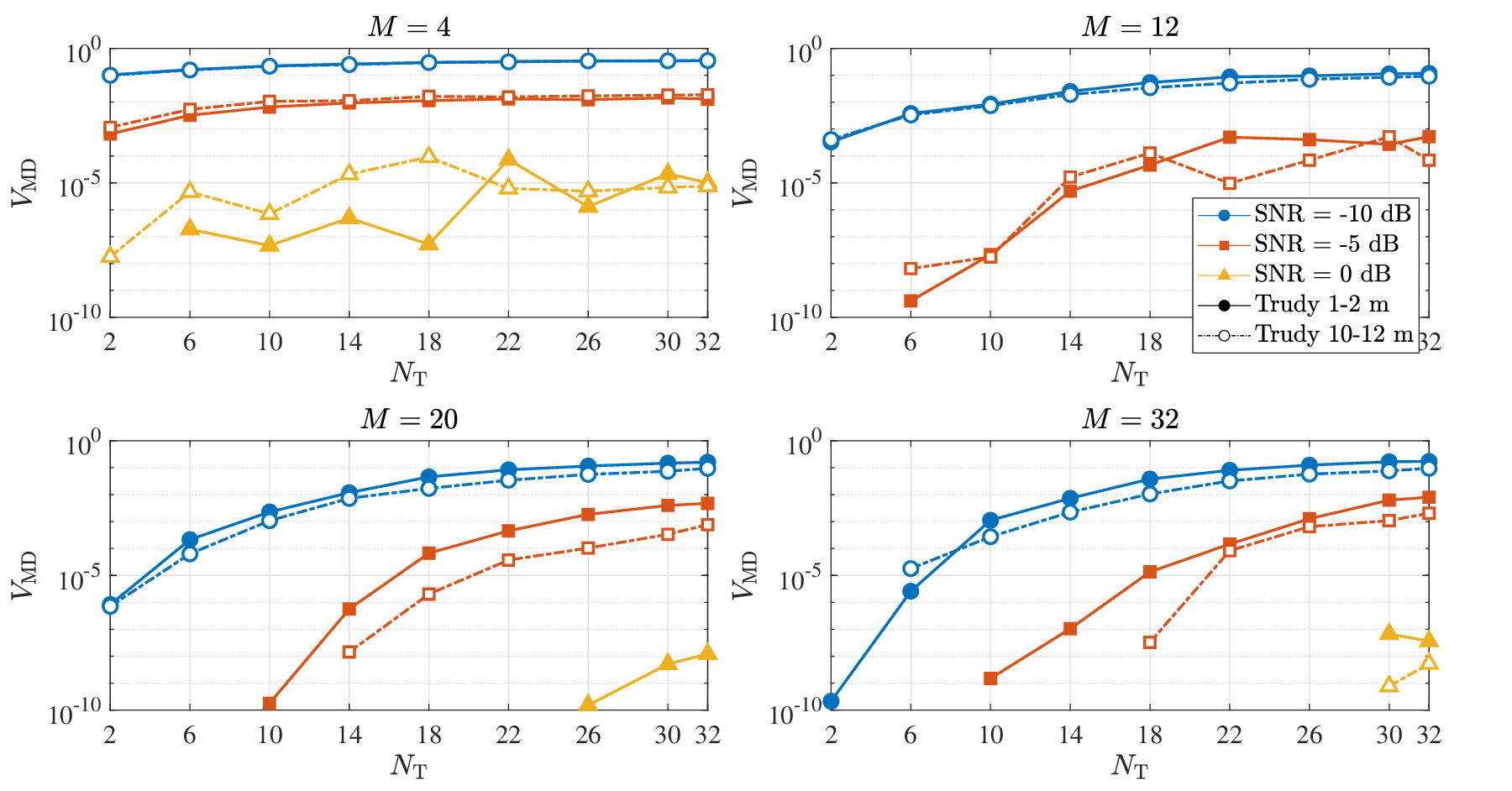}}
    \caption{
    $V_{\mathrm{MD}}$ as a function of the number of antennas at Trudy, $N_{\rm T}$, for $P_{\mathrm{FA}}=10^{-3}$.
    The four panels correspond to different numbers of receive antennas at Bob, $M\in\{4,12,20,32\}$.
    Colors and markers identify the considered SNR values, while filled and empty markers refer to the Trudy--Alice distance ranges $1$--$2$ m and  $10$--$12$ m, respectively.
    }
    \label{fig:pmd_pfa_3}
\end{figure}

Figs.~\ref{fig:pmd_pfa_1} and~\ref{fig:pmd_pfa_3} show the misdetection probability $V_{\mathrm{MD}}$ as a function of $N_{\rm T}$ for $P_{\mathrm{FA}}=10^{-1}$ and $P_{\mathrm{FA}}=10^{-3}$, respectively. In each figure, the four panels correspond to different values of $M$; colors and markers identify the SNR, and filled and empty markers refer to the $1$--$2$~m and $10$--$12$~m intervals, respectively.  
We can observe that $V_{\mathrm{MD}}$ increases with $N_{\rm T}$ without reaching $V_{\mathrm{MD}} = 1 - P_{\mathrm{FA}}$, and remains well below it even for $N_{\rm T}=M=32$. Since Trudy does not know Alice's channel, additional antennas allow her to reduce, but not to cancel, the mismatch between the two cascaded channels.
A strong sensitivity to the SNR is also observed: each SNR increment reduces $V_{\mathrm{MD}}$ by one or more orders of magnitude, and the ordering across SNR values is preserved for every $(M, N_{\rm T})$ pair, since a better channel estimate allows Bob to detect the irreducible mismatch between the two cascaded channels. The threshold behavior in $N_{\rm T}$ is sharp, and the minimum number of antennas required for a non-negligible $V_{\mathrm{MD}}$ grows visibly with $M$. The non-monotonic behavior of the curves with the smallest values of $V_{\mathrm{MD}}$ is due to the fact that, in that regime, the average is dominated by a few channel realizations, rather than to a genuine dependence on $N_{\rm T}$. 
Moreover, the proximity of Trudy to Alice increases $V_{\mathrm{MD}}$, and the gap between the $1$--$2$~m and $10$--$12$~m intervals is wide and persists over a broad range of $N_{\rm T}$, particularly for $M=20$ and $M=32$. Finally, tightening the constraint on the target false-alarm probability, from $P_{\mathrm{FA}} = 10^{-1}$ to $P_{\mathrm{FA}} = 10^{-3}$, shifts all curves upwards, by roughly one to two orders of magnitude for $\mathrm{SNR}\ge -5$~dB, without altering their shape, confirming that the knowledge available to the players and the ratio $N_{\rm T}/M$ govern the qualitative behavior, while $P_{\mathrm{FA}}$ only sets the operating point.
Trudy's codebook also includes, for each of her channel types, a configuration-agnostic precoder. Since in this scenario the dependence between $\bm F$ and $\bm T$ is not described by $\rho$, the estimate~\eqref{eq:Fhat_pure} cannot be used; Trudy instead targets the average $\bar{\bm F}=\frac{1}{N_{\mathcal F}}\sum_{k=1}^{N_{\mathcal F}}\bm F_k$ of Alice's channel types, which coincides with $\mathbb E[\bm F\mid\bm T_\ell]$ since the prior over the types is uniform and factorizes. The precoder is
\begin{equation}
\bm Q_{\mathrm{agn},\ell}=\arg\min_{\bm Q}\sum_{i=1}^{N_\Omega}\big\|\bm G\bm\Omega_i(\bar{\bm F}-\bm T_\ell\bm Q)\big\|_F^2 ,
\end{equation}
which, unlike~\eqref{eq:Qpure}, accounts for the RIS–Bob channel, known to Trudy, by weighting the mismatch uniformly over Bob's codebook.
This inclusion increases $V_{\rm MD}$ by a factor of up to about $2$--$3$ for $M=4$, whereas for $M\ge 20$ its effect is negligible, since Bob's array resolves the residual mismatch. The gain is similar for the two distance intervals, because this precoder exploits the knowledge of Alice's channel statistics rather than the spatial proximity of Trudy to Alice.

\section{Conclusions}
\label{sec:concs}

In Part II of this two-part paper, we investigated RIS-assisted \ac{cr-pla} under partial channel knowledge, where the verifier and the attacker each observe only their own channel realization and rely on statistical knowledge of their opponent's. We formulated the joint design of the RIS configuration policy and the attacker's precoder as a finite-action Bayesian game, in which both players choose mutual \ac{br}s under their respective incomplete information, and characterized the resulting Bayesian Nash equilibrium through the indifference principle. This equilibrium determines Bob's optimal type-dependent randomized RIS policy, Trudy's optimal type-dependent precoding strategy, and the corresponding equilibrium misdetection probability.

Numerical results on geometric multipath and 3GPP TR~38.901 UMi-NLOS channels showed that the equilibrium misdetection probability is governed by the attacker's spatial resources, the correlation between the legitimate and attacker channels, and the SNR. In the geometric scenario, a large receive array and a rich RIS--Bob channel effectively counter a single-antenna attacker, whereas a multi-antenna attacker with a correlated channel approaches the indistinguishability limit. In the 3GPP scenario, the misdetection probability grows with the number of attacker antennas without saturating (nor achieving $1-P_{\textrm{FA}}$) and is highly sensitive to the SNR and to the attacker--user distance. The proposed equilibrium strategy consistently outperforms the considered baselines, and the RIS codebook design induces a clear rate--security tradeoff.

\balance
\bibliographystyle{IEEEtran}
\bibliography{biblio2}

\end{document}